\documentclass[journal]{IEEEtran}
\usepackage{lipsum}
\usepackage[utf8]{inputenc}
\usepackage{amsmath,amsfonts,amssymb,bm}
\usepackage{graphicx}
\usepackage{subcaption}
\usepackage{caption}
\usepackage{xcolor}
\usepackage{mathtools}
\usepackage{comment}
\usepackage{textcomp}

\usepackage{multirow}
\usepackage{epstopdf}
\usepackage{cite,soul}
\usepackage{arydshln}
\usepackage{flushend}
\usepackage{blkarray, multirow}
\usepackage{arydshln}
\usepackage{algorithm}
\usepackage{algpseudocode}
\usepackage{textcomp}
\usepackage{verbatim}
\usepackage{booktabs}
\usepackage{placeins}
\usepackage{tikz}
\usetikzlibrary{shapes,arrows,positioning,calc}
\usepackage{makecell}

\newcommand{\bi}{\begin{itemize}}
\newcommand{\ei}{\end{itemize}}

\newcommand{\be}{\begin{enumerate}}
\newcommand{\ee}{\end{enumerate}}
\newcommand{\bd}{\begin{description}}
\newcommand{\ed}{\end{description}}
\newcommand{\bc}{\begin{center}}
\newcommand{\ec}{\end{center}}
\newcommand{\bt}{\begin{tabbing}}
\newcommand{\et}{\end{tabbing}}
\newcommand{\bfig}{\begin{figure}}
\newcommand{\efig}{\end{figure}}
\newcommand{\beq}{\begin{equation}}
\newcommand{\beqarr}{\begin{eqnarray}}
\newcommand{\beqarrn}{\begin{eqnarray*}}
\newcommand{\eeq}{\end{equation}}
\newcommand{\eeqarr}{\end{eqnarray}}
\newcommand{\eeqarrn}{\end{eqnarray*}}
\newcommand{\bflr}{\begin{flushright}\vspace{-0.2in}}
\newcommand{\eflr}{\end{flushright}}
\newcommand{\bsub}{\begin{subequations}}
\newcommand{\esub}{\end{subequations}}
\newcommand{\barr}{\begin{array}}
\newcommand{\earr}{\end{array}}
\newcommand{\nn}{\nonumber}

\def\undb#1{\mbox{\bf{#1}}}

\def\dn{\stackrel{\scriptscriptstyle \triangle}{=}}

\def\BibTeX{{\rm B\kern-.05em{\sc i\kern-.025em b}\kern-.08em
		T\kern-.1667em\lower.7ex\hbox{E}\kern-.125emX}}

\begin{document}
\title{Compact Reconfigurable Intelligent Surface with Phase-Gradient Coded Beam Steering and Controlled Substrate Loss}
\author{Mahendra Kheti, \IEEEmembership{Student Member, IEEE}, Debapratim Ghosh, \IEEEmembership{Senior Member, IEEE}, \\ and Soumya P. Dash, \IEEEmembership{Senior Member, IEEE}
\thanks{An Indian patent bearing application number 202631022577 on this work was filed on 25 February 2026.}
\thanks{The authors are with the Department of Electronics and Communication Engineering, School of Electrical and Computer Sciences, Indian Institute of Technology Bhubaneswar, Argul, Khordha, Odisha 752050 India, e-mail: \{a25ec09009, debapratim, spdash\}@iitbbs.ac.in}
}

\maketitle

\begin{abstract}
This paper presents a 1-bit reconfigurable intelligent surface (RIS) fabricated using a three-layer structure. It employs a manual layer stackup incorporating an optimal air gap to reduce the effective dielectric losses while using a low-cost FR4 substrate. The new design of the unit cells of the proposed RIS is outlined, with each unit cell featuring a PIN-diode-based, compact, simplified biasing network that simplifies the control circuit while maintaining distinct $\boldsymbol{0^\circ/180^\circ \pm 20^\circ}$ phase states between ON/OFF conditions. The designed RIS is in the form of a $\boldsymbol{10\times10}$ array with a compact size of $\boldsymbol{2.9\lambda_g \times 2.9\lambda_g}$. Additionally, a phase-gradient coding scheme is presented and utilized that achieves measured beam steering up to $\boldsymbol{\pm30^\circ}$ in both anechoic and noisy environments. Controlled and driven by an Arduino-cum-digital interface, the proposed RIS exhibits measured reflected wave gain enhancement of about 9\,dB over an incident wave angular range of $\boldsymbol{\pm 30^\circ}$. Furthermore, the design is also experimentally validated by transmitting quadrature phase-shift keying-modulated symbols via the RIS-assisted wireless channel. The proposed RIS works for the range 3.38--3.67\,GHz (8.3\%), and is suitable for deployment for the 5G n78 \mbox{band (3.5\,GHz).}

\end{abstract}


\vspace{-0.05cm}
\section{Introduction}
\label{sec:introduction}

\IEEEPARstart{T}{he} development of fifth generation (5G) and beyond-5G (B5G) wireless networks is expected to support enhanced wireless bandwidth, ultra-reliable and low-latency communication, and massive machine-type communications \cite{b1}. Some relevant challenges include improvement of spectral efficiency and network coverage arising due to unfavorable propagation conditions, blockage, and severe multipath fading, particularly in dense urban and indoor environments \cite{b2}. To overcome these challenges, Reconfigurable Intelligent Surfaces (RISs), which enable programmable control to adjust the reflected beam characteristics, have been identified as promising enablers for 5G and B5G systems \cite{b3, b4}. Consequently, low-cost RIS hardware implementations for bands above 3\,GHz are essential for practical deployment.

The practical implementation of an RIS involves several interdependent design aspects, including unit-cell geometry, phase-control mechanisms, substrate selection, biasing network implementation, and associated control circuitry. In theory, a unit cell of an RIS should offer a continuous reflection phase range up to $360^\circ$ \cite{29,35}. However, such continuous-phase implementations require highly complex biasing and control circuits and often suffer from indistinguishable phase states. To address these issues, discrete-phase RIS architectures have been reported using tuning devices such as varactors, PIN diodes, MEMS, etc., where a finite number of phase states is utilized to reduce hardware complexity while preserving acceptable beam steering performance \cite{b6,b7,b9,b10,b15, b18, 24}.
However, in these papers, several design aspects remain insufficiently addressed, particularly the aperture size reduction of the overall RIS. Furthermore, many reported prototypes rely on specialized low-loss substrates, which increases overall cost. Some previously reported RISs focus on improving beam steering capabilities using programmable metasurfaces and PIN-diode-based reflective structures \cite{b3}, network-based modeling \cite{34}, and digitally coded metasurfaces \cite{26}. While programmable metasurfaces using multi-bit PIN-diode-loaded unit cells have demonstrated beam steering and polarization control \cite{b3,b7}, these configurations are typically large in size and need extremely complex control circuits. In addition, aspects such as optimized via placement and the design of compact biasing layers for reliable PIN-diode switching have not been adequately addressed.

To address these challenges, this work proposes an RIS configuration that focuses on reducing the dielectric loss in a low cost substrate, an approach that has not been adequately explored thus far. Our proposed approach starts by designing a new unit cell whose reflected phase performance is optimized on multiple fronts, viz., a controlled introduction of an air gap between PCB layers, and strategic placement of vias that result in the desired phase performance. This also results in compact unit cells, thereby creating a dense and overall compact array. A comparatively simplified biasing network and control circuit are developed and utilized. Most importantly, to obtain the desired beam steering, a phase gradient-based scheme is also developed.
In summary, this work makes the following specific contributions:
\begin{itemize}
\item A new RIS unit cell of size $17\text{mm} \times 17\text{mm}$ $(0.198\lambda_g \times 0.198\lambda_g)$ designed using a controlled air gap, Figure of Merit (FoM), and optimum via-placement approach. 

\item A new and simplified biasing network using PIN diodes and digital control architecture using standard digital ICs and a microcontroller. 

\item A $10 \times 10$ RIS prototype with area $255\text{\,mm}\times 255\text{\,mm}$ with measured beam steering up to $\pm 30^\circ$ obtained by using a phase gradient coded approach, operating over 3.38--3.67\,GHz ($8.3\%$).

\end{itemize}

The rest of the paper is organized as follows. Section \ref{suc} presents the RIS unit-cell design, including geometry, layered structure, phase sensitivity, and via placement under Floquet excitation. Section \ref{stm} describes the theoretical modeling of the planar reflection-mode RIS, followed by the formulation of the far-field scattered fields. Section \ref{ssm} details the experimental setup and measurement procedure, and Section \ref{svm} presents the experimental verification and results. Section \ref{gnuradio} demonstrates the effect of RIS on the transmission of quadrature phase-shift keying (QPSK) modulated signals, and Section \ref{sc} concludes the paper.
\vspace{-0.05cm}
\section{Proposed Unit Cell Structure and Modeling}
This section presents the detailed design of the proposed 1-bit RIS unit cell, including its geometric configuration, multilayer structure, and the PIN-diode-based switching mechanism. The electromagnetic response is analyzed under Floquet port excitation to characterize the reflection phase and amplitude performance. 
\label{suc}

\subsection{Geometry and Layered Structure}
\begin{figure}[!t]
    \centering
    \includegraphics[width=1\linewidth]{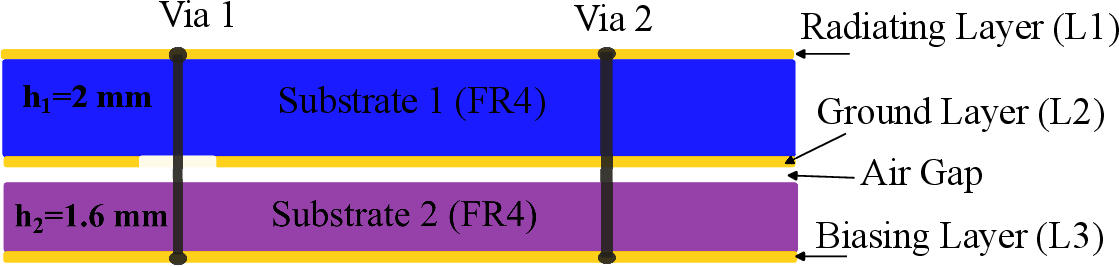}
    \caption{Cross-sectional view of the proposed RIS unit cell illustrating various layers.}
    \label{fig:overal}
\end{figure}
A unit cell of the proposed 1-bit RIS is designed based on a three-layer structure, consisting of a radiating layer (L1), a ground layer (L2), and a dedicated biasing layer (L3), as illustrated in Fig.~\ref{fig:overal}. The design incorporates two FR4 substrates, namely Substrate 1 with thickness $h_1 = 2$\,mm, placed between the radiating and the ground layer, and Substrate 2 with thickness $h_ 2 = 1.6$\,mm. FR4 substrates have $\varepsilon_\text{r}\approx 4.4$ and $\tan \delta \approx 0.02$. Moreover, an air gap, outlined later in this section as a part of further analysis (c.f. (\ref{h})), introduced between the ground layer and Substrate 2, aims to reduce the effective dielectric constant of the RIS's unit cell, thus reducing its overall RF losses. From (\ref{eq1}) and (\ref{eff_height}), the optimal values of $h_1$ and $h_2$ are obtained while ensuring a compact structure.

\begin{figure}[!t]
\centering

\subfloat[]{\includegraphics[width=0.45\columnwidth]{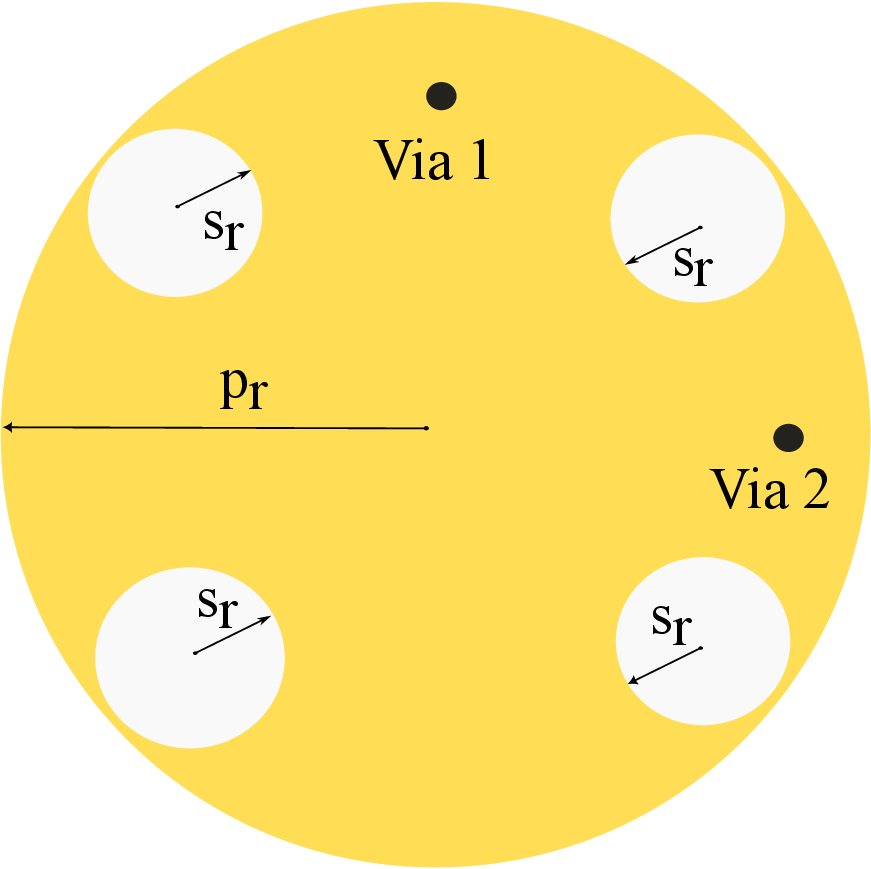}
\label{rl}}\hfill
\subfloat[]{\includegraphics[width=0.45\columnwidth]{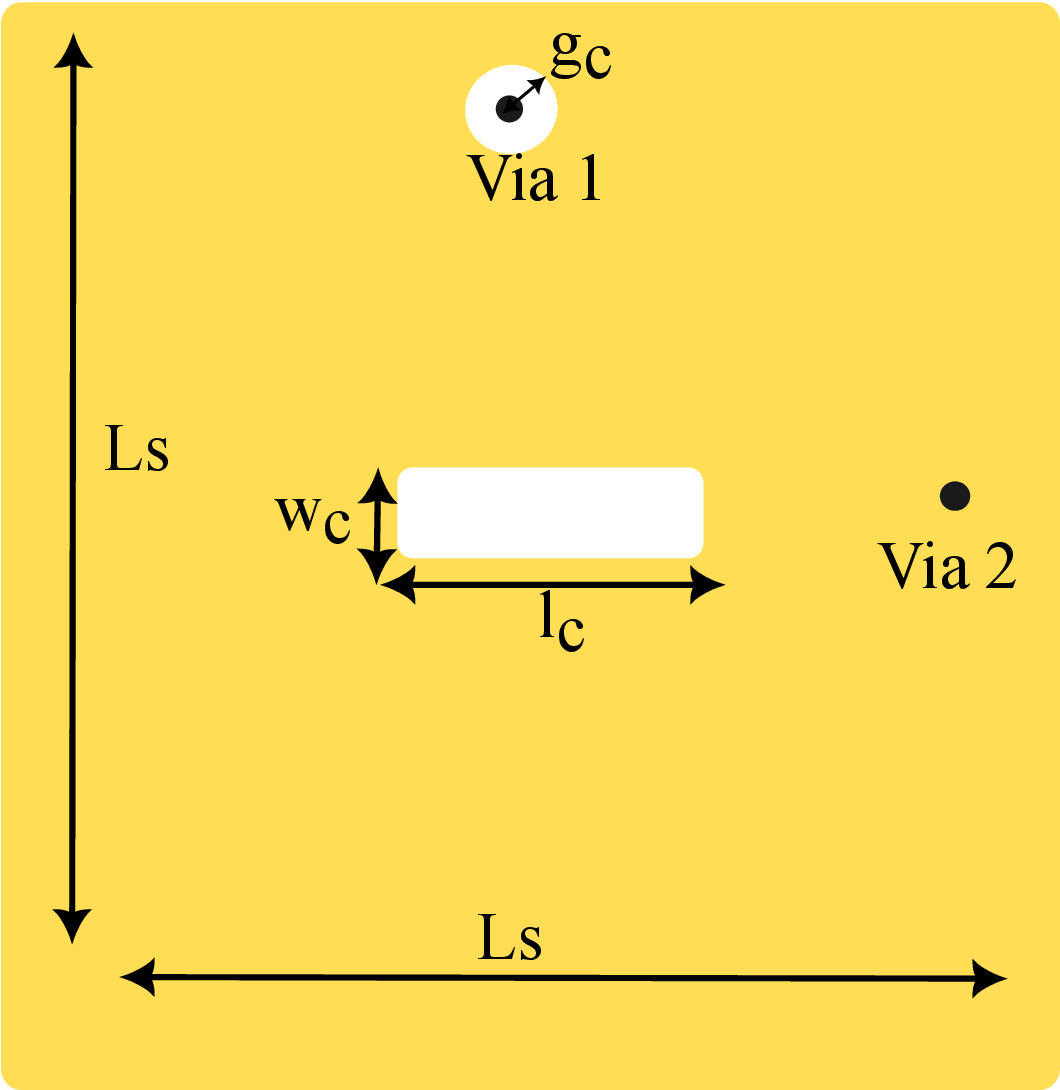}
\label{gnd}}\\

\subfloat[]{\includegraphics[width=0.49\columnwidth]{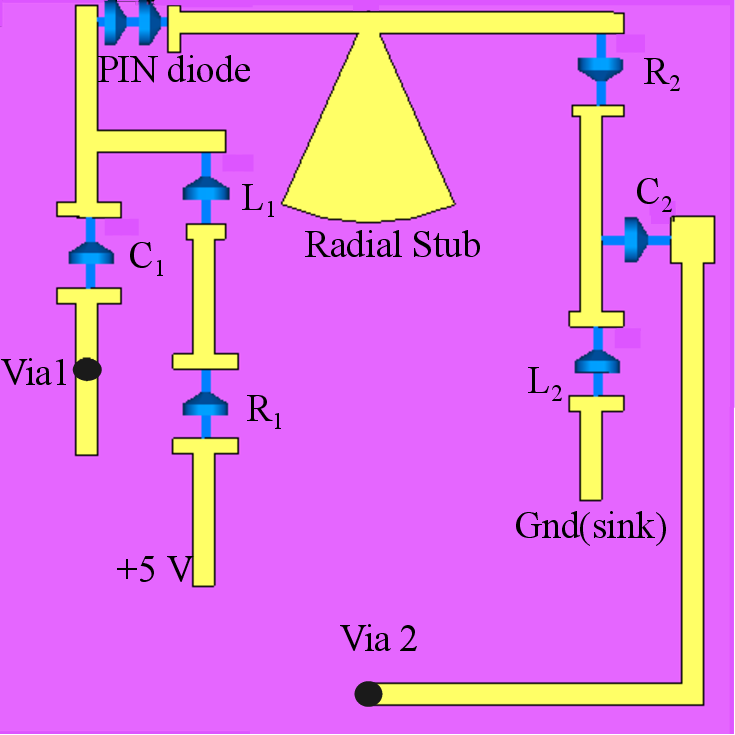}
\label{bias}}\hfill
\begin{minipage}{0.45\columnwidth}
\centering
\vspace{-4cm}
\subfloat[]{\includegraphics[width=0.7\linewidth]{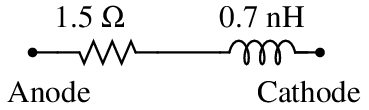}
\label{on}} \\
\vspace{1mm}
\subfloat[]{\includegraphics[width=\linewidth]{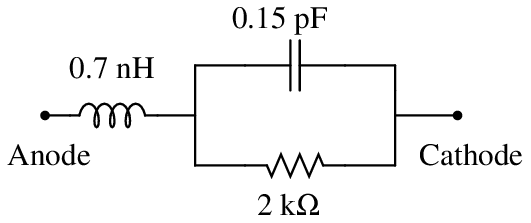}\label{off}}
\end{minipage}

\caption{(a) Radiating layer (b) ground layer and (c) bias layer of the RIS unit cell. Equivalent circuits of the PIN diode (Skyworks SMP1345-079LF) in (d) ON state and (e) \mbox{OFF state.}}
\label{fig:matrix}
\end{figure}

The radiating (top) layer consists of a circular metallic patch with radius $p_\text{r}=8.49$\,mm. It has four identical circular slots of radius $s_\text{r}=1.6$\,mm that are placed symmetrically, as shown in Fig.~\ref{rl}. The circular geometry provides rotational symmetry, ensuring polarization-independent reflection and stable phase response for varying incidence angles \cite{b12,b14}. The bottom layer of Substrate 1 is a ground plane (size $L_s=17$\,mm, $g_c=0.2$\,mm) having a rectangular slot ($w_c=0.75$\,mm, $l_c=2$\,mm), as shown in Fig.~\ref{gnd}. This slotted ground controls the internal field distribution, improving impedance matching and $S_{11}$ performance. A vertical air gap \cite{b11} is introduced to enhance phase control, followed by Substrate 2 with the biasing layer on the bottom, manually stacked up. The unit cell has a size of $17\times 17\,\text{mm}^2$ $(0.198\lambda_g \times 0.198\lambda_g)$ at the center frequency of 3.5\,GHz.

The unit cell forms a multilayer resonant cavity. Compared to a conventional single-substrate microstrip antenna, the introduced air gap modifies the internal field distribution and influences the resonant behaviour. Due to the small air gap height, only transverse magnetic (TM) modes are assumed to exist \cite{b10,b11}. Since two media are present along the vertical direction, the longitudinal electric field varies across the dielectric–air interface. For electrically thin substrate and air-gap layers, this variation can be approximated using an effective dielectric constant \cite{b14,27}. The effective dielectric constant of the two-layer cavity mainly depends on the substrate thicknesses $h_\mathrm{1}, h_\mathrm{2}$, the air gap height $h_\mathrm{air}$, and the absolute $\varepsilon_\mathrm{r}$=$\varepsilon_\mathrm{air}$$\varepsilon_\mathrm{sub}$, and can be approximated as \cite{b18}
\begin{equation}
\varepsilon_{\mathrm{eff}}
= \frac{\varepsilon_{\mathrm{r}} \left(h_{1}+h_{2}+h_{\mathrm{air}}\right)}
{\left(h_{1}+h_{2}\right)\varepsilon_{\mathrm{sub}}
+\varepsilon_{\mathrm{air}}h_{\mathrm{air}}} \, ,
\label{eq1}
\end{equation}
where $\varepsilon_\mathrm{sub}$ and $\varepsilon_\mathrm{air}$ are the dielectric constants of the substrate and air gap, respectively. Since $\varepsilon_\mathrm{air} < \varepsilon_\mathrm{sub}$, it is to be noted from \eqref{eq1} that increasing $h_\mathrm{air}$ reduces the effective dielectric constant, causing an upward shift in the resonant frequency. This behaviour provides an effective approach for resonance tuning and bandwidth enhancement \cite{b14,b15}. To obtain the required $\varepsilon_{\mathrm{eff}}$, the substrate thicknesses $h_1$ and $h_2$ are selected based on commercially available low-cost FR4 substrates, while the air-gap thickness $h_{\mathrm{air}}$ is determined using (\ref{h}). In the proposed RIS, this mechanism enables controlled resonance tuning while maintaining electromagnetic isolation between the radiating and biasing layers. Considering all layers and the effective dielectric constant due to the air gap, the electrical thickness of the entire unit cell is computed as
\begin{equation}
h_{\mathrm{eff}} = \frac{h}{\lambda_0 \sqrt{\varepsilon_{\mathrm{eff}}}} \, ,
\label{eff_height}
\end{equation}
where $h$ is the total physical thickness of the unit cell and $\lambda_0$ is the free-space wavelength.

Typically, FR4 substrates are less preferred in RIS implementations above 3\,GHz due to their high solid state losses. However, we demonstrate theoretically that the sensitivity of substrate loss is not solely determined by the intrinsic loss tangent of the material. Instead, it is also influenced by the distribution of electromagnetic field energy within the substrate, since the resulting dielectric power dissipation depends on both material loss parameters and the electric field intensity.
The dielectric loss power in a material region is given by \cite{b19}
\begin{equation}
P_\text{d} = \frac{\omega \varepsilon_0 \varepsilon_\text{r} \tan\delta}{2} \int_{V_\text{d}} |E(\mathbf{r})|^2 \, \text{d} V \, ,
\end{equation}
where $\omega$ is the angular frequency, $\varepsilon_0$ is the dielectric constant of free space, $\tan \delta=0.02$ is the material standard value, provided by the manufacturer, and $E$ is the electric field within the volume of the lossy dielectric region $V_\text{d}$. To capture the influence of FR4 on the overall loss, we define a \emph{loss participation ratio} (LPR), extended from \cite{b19}, as
\begin{equation}
\mathrm{LPR} \propto
\frac{\int_{V_{\mathrm{FR4}}} |E(\mathbf{r})|^2 \, \text{d} V}
{\int_{V_{\mathrm{total}}} |E(\mathbf{r})|^2 \, \text{d} V} \, ,
\end{equation}
where $V_{\mathrm{FR4}}$ and $V_{\mathrm{total}}$ represent the volume in a single FR4 layer and the entire unit-cell volume, with the complete layer stackup, respectively. An effective loss tangent can then be expressed as
\begin{equation}
\tan\delta_{\mathrm{eff}} = \mathrm{LPR} \tan\delta_{\mathrm{FR4}} \, .
\label{eff_tan}
\end{equation}
Hence, due to the air-gap-assisted field redistribution, $\mathrm{LPR} \ll 1$, resulting in a significantly reduced effective loss tangent despite the use of FR4 substrates. This LPR-based analysis leads to an optimized structure in terms of reduced losses and compact design. As noted from (\ref{eff_tan}), the desirable magnitude and phase responses of ($S_{11}$) can be achieved.

To assess the phase robustness against loss, we define a phase-to-amplitude FoM as
\begin{equation}
\mathrm{FoM}_{\phi} \dn \frac{\Delta \phi}{\Delta |S_{11}|} \, ,
\end{equation}
where $\Delta \phi$ represents the achievable reflection phase tuning range of the proposed design, while $\Delta |S_{11}|$ denotes the corresponding achievable average reflection magnitude from the ON and OFF states. A high $\mathrm{FoM}_{\phi}$ confirms that flexible phase control is effectively maintained, even in the presence of a lossy substrate. Since $\varepsilon_{\mathrm{eff}}$ and $h_{\mathrm{eff}}$ depend on the height of the air gap ($h_{\mathrm{air}}$), it becomes a key design parameter. The optimal $h_{\mathrm{air}}$ height is obtained by maximizing the FoM as
\begin{equation}
h_{\mathrm{air}}^{\mathrm{opt}}=\arg\max_{h_{\mathrm{air}}}\{\mathrm{FoM}_{\phi}\} \, .
\label{h}
\end{equation}

During the simulation studies, the values of ${h_{air}}$ were varied within a practical range of 0.05\,mm to 1.5\,mm, and the optimal performance was observed at a finely tuned  $h_\text{air}$ of 0.5\,mm, which provides the best trade-off between phase control and minimal RF losses. Increasing $h_{\mathrm{air}}$ initially improves $\mathrm{FoM}_{\phi}$ due to reduced electric-field confinement in FR4 and enhanced phase coverage. Beyond a critical height, further improvement becomes marginal, while the overall thickness of the RIS increases. Accordingly, $h_{\mathrm{air}}^{\mathrm{opt}}=0.5$\,mm is selected as the air-gap height that provides near-maximum $\mathrm{FoM}_{\phi}$, ensuring low loss and a compact structure.

\begin{figure}[!t]
    \centering
    \includegraphics[width=\linewidth]{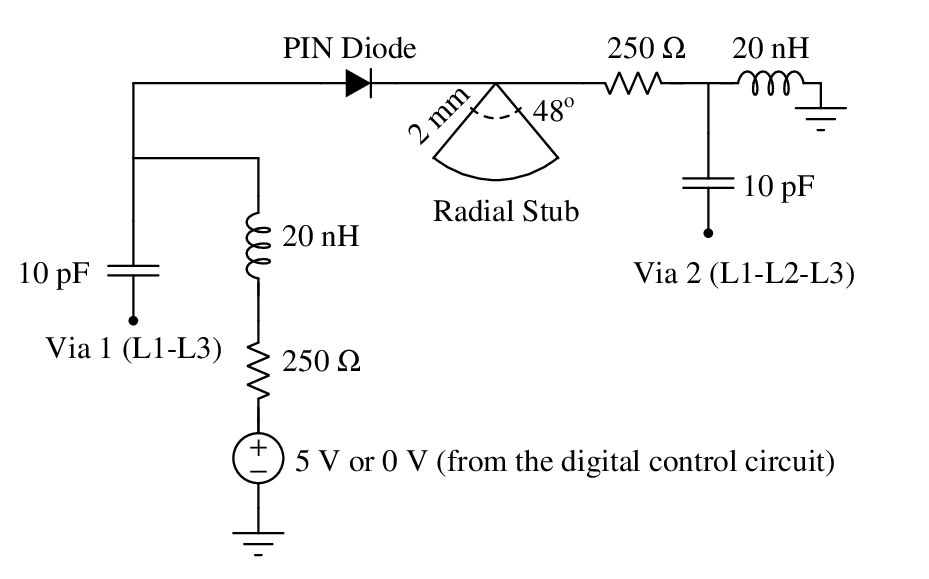}
    \caption{Equivalent circuit model of the proposed biasing layer, including the PIN diode and associated passive components.}
    \label{eqv}
\end{figure}
To enhance the reconfigurability and operational reliability of the unit cell, a biasing layer has been designed that integrates a dedicated PIN diode (Skyworks SMP1345-079LF) \cite{pindiode}, allowing dynamic switching between ON and OFF states with a controlled 10\,mA bias current driven by a DC line (5\,V or 0\,V) from the control circuit, providing precise control over the switching states. The equivalent circuit representations of the ON and OFF states are shown in Fig.~\ref{on} and Fig.~\ref{off}. This proposed biasing network design implements dual-path DC/RF isolation using discrete reactive elements. Specifically, high-bandwidth series inductors (chokes) are employed to block high-frequency RF signals from propagating into the DC supply path, while capacitors prevent DC from entering the RF path. This configuration ensures effective separation of the RF and DC path, minimizes parasitic coupling, and maintains broadband impedance matching. A radial stub with a precisely optimized length of 2\,mm and angle of 48$^\circ$ is incorporated immediately after the diode to enhance RF isolation further and provide a broadband RF open circuit, contributing to improved operational bandwidth. This simplified biasing network is successfully used and is important when compared to conventional multi-stage circuits reported in the literature. This network is designed to maintain uniform bias distribution across the array while consuming minimal DC power of approximately 0.05\,W per unit cell. The combined effect of the diode, inductive and capacitive isolation elements, and the radial stub is captured in the equivalent circuit model of the biasing layer as shown in Fig.~\ref{eqv}. This design represents integrated DC/RF biasing in the RIS, achieving high-performance reconfigurability, broadband operation, and enhanced reliability while minimizing DC power consumption.
\subsection{Phase Sensitivity and Via Placement Under Floquet Port Excitation}
The standard designs reported in \cite{29,b19} use vias that are typically placed at electric field minima point(s) to minimize RF disturbances. In this paper, we perform a Floquet-mode analysis of the unit cell to determine the optimized via placement, taking into account the dominant surface current distribution and its effect on the reflection phase. This allows us to identify a location that simultaneously preserves the full reflection coefficient phase range ($360^\circ$) along with high reflection magnitude, demonstrating an application specific design methodology for RIS unit cells.

The RIS unit cell is analyzed under periodic boundary conditions in the $x$–$y$ plane and excited using Floquet ports to model an infinite array. A fundamental Floquet mode, assuming linear polarization is applied, where the passive surface current density $\mathbf{J}_s(\mathbf{r})$ is determined by the metallization geometry and slot-induced surface impedance distribution. Under Floquet excitation, the surface current on the unit cell can be expressed as a spatial harmonic expansion \cite{32} as

\begin{equation}
\mathbf{J}_s(\mathbf{r}) =
\sum_{m,n} \mathbf{J}_{mn}
e^{- \jmath \mathbf{k}_{mn} \cdot \mathbf{r} } \, ,   (m,n) \in \mathbb{Z} \, ,
\label{eq:floquet_current}
\end{equation}
where $\jmath=\sqrt{-1}$, $ \cdot $ outputs the inner-product, $(m,n)$ denote the Floquet modes with $(0,0)$ corresponding to the fundamental mode governing the reflected field \cite{b20,32}.

Prior to introducing the switching element, i.e., the PIN diode, the passive unit cell of the RIS is used to identify regions with maximum surface current densities. The surface current satisfies the continuity condition $\nabla_s \cdot \mathbf{J}_s = 0$, over the metallized region, while the etched slots impose high surface impedance boundaries \cite{b21}.
As a result, the dominant current is constrained to flow through narrow conductive regions connecting adjacent impedance discontinuities.
When a metallic via with impedance $Z_{via}$ is introduced at location $\mathbf{r}$, it acts as a localized shunt perturbation to the surface current associated with the fundamental Floquet mode. Using first-order perturbation analysis\cite{32}, the resulting phase variation of the fundamental reflection coefficient can be expressed as
\begin{equation}
\Delta \angle \Gamma_{00}
\propto
\left|\mathbf{J}_{00}(\mathbf{r})\right|^2
\, \Delta Z_{via} \, ,
\label{eq:phase_perturbation}
\end{equation}
where $\mathbf{J}_{00}(\mathbf{r})$ denotes the surface current corresponding to the $(0,0)$ Floquet mode \cite{b20}.
Further, \eqref{eq:phase_perturbation} indicates that the phase sensitivity is maximized when the via is placed at a location where the passive fundamental-mode surface current density is maximum. Accordingly, the optimal via position is determined as
\begin{equation}
\mathbf{r}_{via}^{opt}
=
\arg\max_{\mathbf{r} \in S}
\left|\mathbf{J}_{00}(\mathbf{r})\right| \, ,
\label{eq:via_location}
\end{equation}
where $S$ denotes the metallized area of the unit cell. The via is placed at the centroid of this region to ensure stable coupling to the dominant surface current path. Full-wave simulations using periodic boundaries and Floquet ports are subsequently employed to verify the predicted current concentration and to quantify the achievable phase-tuning range.

\begin{figure}[!t]
\centering
\includegraphics[width=\linewidth,trim=60 20 90 20,clip]{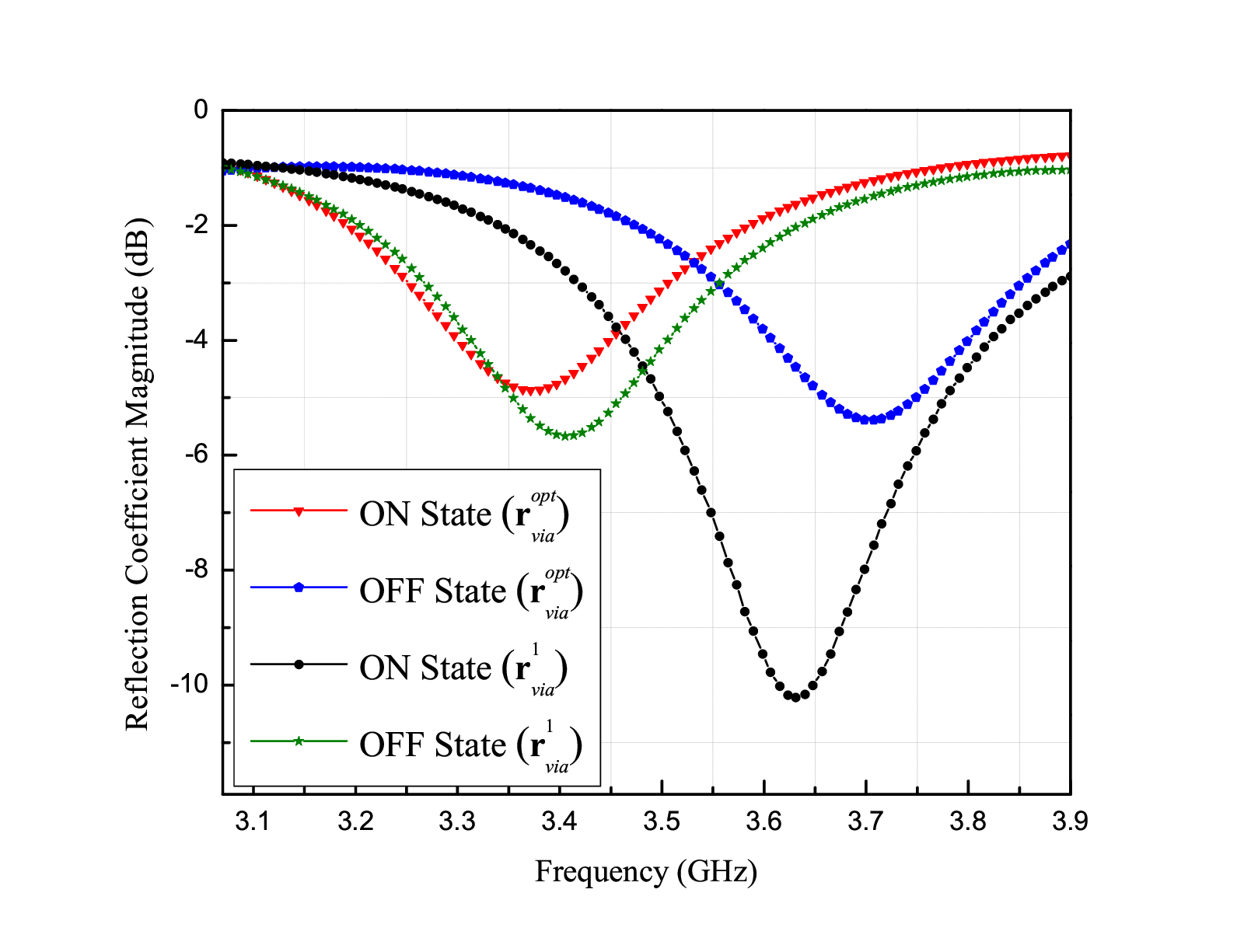}
\caption{Simulated reflection coefficient of the RIS unit cell under different operating conditions.}
\label{fig:ucr}
\end{figure}

\begin{figure}[!t]
    \centering
    \includegraphics[width=\linewidth,trim=45 20 90 20,clip]{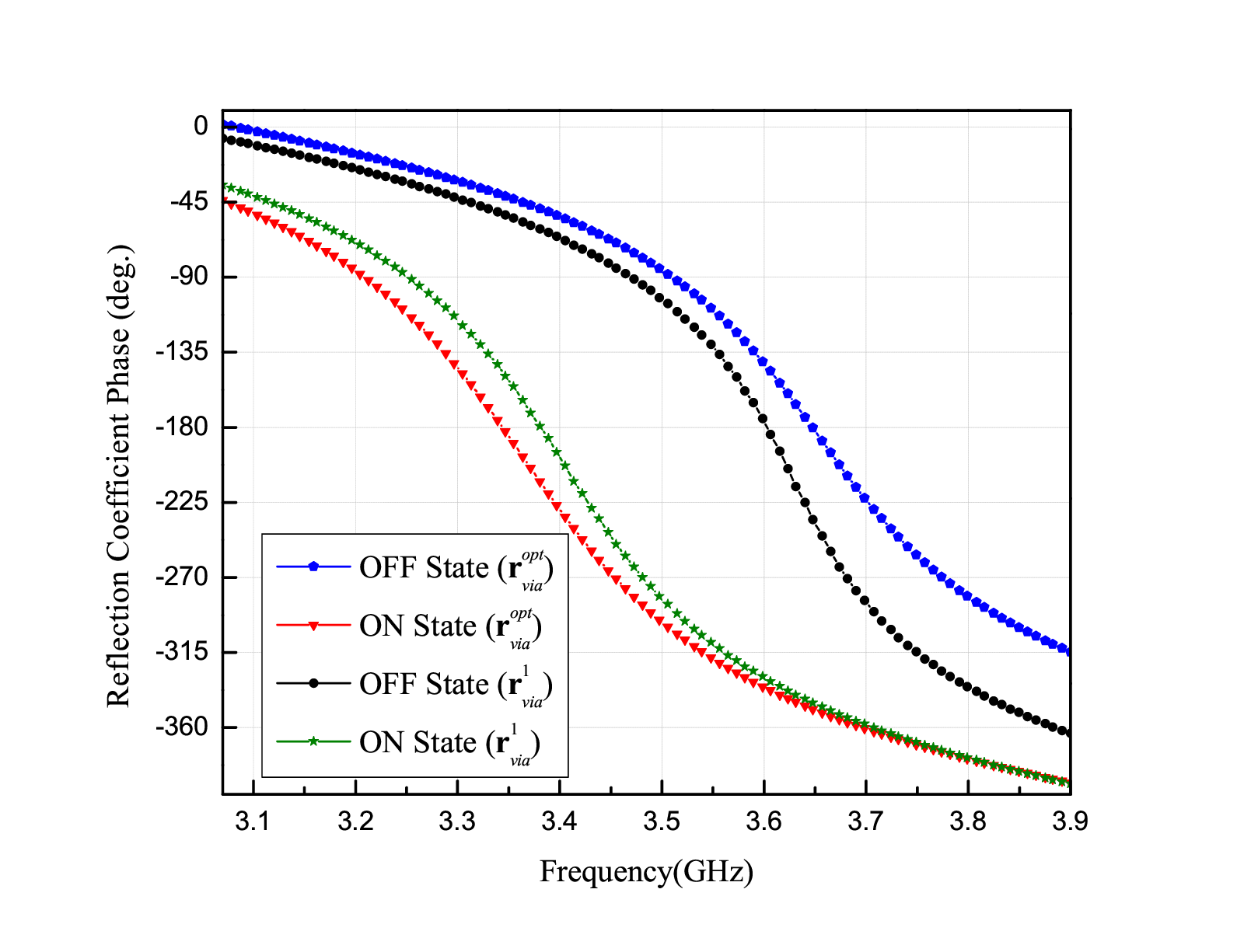}
    \caption{Simulated reflection coefficient phase response of the RIS unit cell under different operating conditions.}
    \label{fig:placeholder}
    \label{fig:result}
\end{figure}

The simulated $\lvert S_{11} \rvert$, shown in Fig.\ref{fig:ucr} indicate the optimized configuration, $\undb{r}_{via}^{opt}$. Ideally, $|S_{11}|$ should approach 0\,dB with a full $360^\circ$ reflection phase range; however, in practice, a trade-off arises due to material properties. The optimized proposed design achieves the best performance, with $|S_{11}|$ of $-4.9$\,dB in the ON state and $-5.7$\,dB in the OFF state. The corresponding phase responses are presented in Fig.~\ref{fig:result}, highlighting the phase-tuning capability of the proposed unit cell. A maximum phase variation of approximately $320^\circ$ is achieved, demonstrating effective phase control within the operating band. Furthermore, the phase difference between the two reconfigurable ON/OFF states remains within the range of $0^\circ-180^\circ \pm20^\circ$  over the frequency span from 3.38\,GHz to 3.67\,GHz. Moreover, compared to the non-optimized location $\left(\undb{r}_{via}^{1}\right)$, the optimized position $\left(\undb{r}_{via}^{opt} \right)$ achieves a broader bandwidth and an improved reflection coefficient. Thus, the impedance bandwidth spans from 3.38\,GHz to 3.67\,GHz, corresponding to a fractional bandwidth of approximately 8.3\%. Thus, we show here that using a proposed via placement technique utilizing a Floquet port excitation, we are able to obtain the optimized via positions in the unit cell, and this is corroborated by the simulated $S$-parameter magnitude and phase responses in both ON and OFF states.
\vspace{-0.05cm}
\section{Theoretical Modeling of Planar Reflection-Mode RIS}
\label{stm}
\begin{figure}[!t]
\centering
\includegraphics[
width=1\linewidth,
height=6cm,
keepaspectratio, trim=0mm 15mm 0mm 0mm,
clip
]{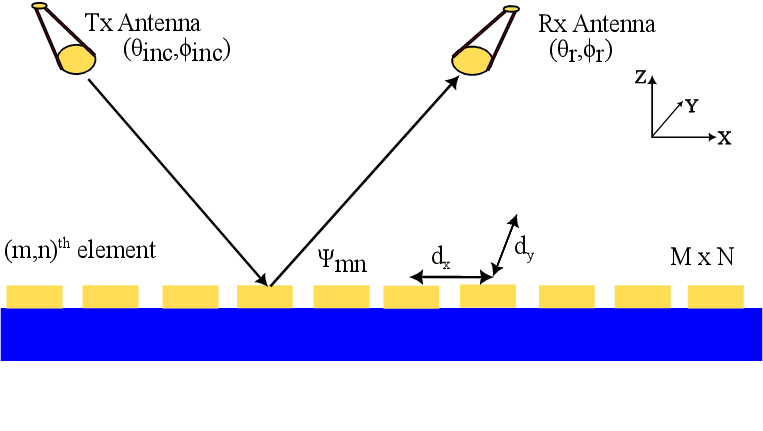}
\caption{Proposed RIS reflective surface showing different quantities.}
\label{fig:rs}
\end{figure}

Following the design of a unit cell of the proposed 1-bit RIS, it is essential to independently control each unit cell across the radiating surface to achieve effective beam steering. Beam manipulation in a reconfigurable structure is fundamentally governed by the spatial phase distribution, which can only be realized when the individual unit cells are selectively tuned as shown in Fig.\ref{fig:rs}.

The RIS is illuminated by a rectangular microstrip patch antenna (MPA), a standard source which is positioned such that the RIS lies in far-field region. Under this condition, the wavefront incident on the RIS aperture can be approximated as a plane wave. We thus consider a planar reflection mode RIS composed of $M \times N$ sub-wavelength unit cells arranged on a flat conducting ground plane. Each unit cell is characterized by a tunable reflection coefficient in the polar form given by
\begin{equation}
\Gamma_{mn} = \alpha_{mn} e^{\jmath \psi_{mn}} \, , m=1 , \ldots,M \, , n=1,\ldots,N \, ,
\label{refl}
\end{equation}
where $\alpha_{mn} \leq 1$ represents the reflection amplitude accounting for material losses, and $\psi_{mn}$ is the programmable reflection phase.
The position vector of the $(m, n)^{\text{th}}$ element of the RIS can be expressed as
\begin{align}
& \mathbf{p}_{mn} = x_m \hat{\mathbf{x}} + y_n \hat{\mathbf{y}},
\quad
x_m = (m-1) d_x, \quad y_n = (n-1) d_y \, , \nn \\
& \qquad \qquad \qquad \qquad \quad
m=1 , \ldots,M \, , n=1,\ldots,N \, ,
\label{position}
\end{align}
where $d_x$ and $d_y$ denote the inter-element spacings along the $x$- and $y$-directions, respectively. 

The incident electric field at position $\mathbf{p}_{mn}$ of the RIS can be expressed as
\begin{equation}
\mathbf{E}_{\mathrm{inc}}\left( \mathbf{p}_{mn} \right) 
= E_0 \, e^{- \jmath k_0 \hat{\mathbf{u}}_{\mathrm{inc}} \cdot \mathbf{p}_{mn} } ,
\label{eq:incident_field}
\end{equation}
where $E_0$ denotes the amplitude of the incident electric field, $k_0 = 2\pi/\lambda_g$ is the free-space wavenumber, and $\hat{\mathbf{u}}_{\mathrm{inc}}$ is the unit vector defining the direction of incidence, and is given by $\hat{\mathbf{u}}_{\mathrm{inc}}
= \hat{\mathbf{x}} \sin\theta_{\text{inc}} \cos \phi_{\text{inc}}
+ \hat{\mathbf{y}} \sin\theta_{\text{inc}} \sin \phi_{\text{inc}}
+ \hat{\mathbf{z}} \cos \theta_{\text{inc}} \, ,$
where $\theta_{\text{inc}}$ and $\phi_{\text{inc}}$ represent the elevation and azimuth angles of the incident wave, respectively.

Since the incident wave arrives at an angle to the RIS, the incident phase varies spatially across the RIS aperture which lies in the $x-y$ plane and must be explicitly considered. This results in the phase of the $(m, n)^{\text{th}}$ element of the RIS to be given by $
\psi_{mn}
= - k_0 \left( \mathbf{p}_{mn} \cdot \hat{\mathbf{u}}_{\mathrm{inc}} \right)\, , $ which introduces a linear phase gradient over the surface. This spatial variation directly affects the required reflection phase of each RIS element and needs to be compensated in the beamforming design. To ensure constructive interference in that direction. 
the required phase profile of 
the $(m, n)^{\text{th}}$ element can be expressed as
\begin{equation}
\psi_{mn} = - k_0 \left( x_m \sin \theta_{\text{inc}} \cos \phi_{\text{inc}} 
+ y_n \sin\theta_{\text{inc}} \sin\phi_{\text{inc}}  \right)  \, .
\label{eq:phase_law}
\end{equation}
This phase profile 
establishes a linear phase gradient across the RIS aperture, 
resulting in a reflected plane wave pointing toward 
$(\theta_\text{r},\phi_\text{r})$. 

In practice, RIS elements cannot realize a continuous reflection phase because of hardware constraints. Instead, each element only produces a finite set of discrete phase states. For a 1-bit RIS architecture, the realizable phase values can be $\{0,\pi\}$. Consequently, the desired continuous phase distribution $\psi_{m,n}$ obtained from the beamforming design cannot be directly implemented and must be mapped to the available discrete states.

The reflected wave from the RIS is assumed to travel towards an observation direction $(\theta_\text{r},\phi_\text{r})$ with element pattern $f(\theta_\text{r},\phi_\text{r})$. Under far-field conditions, the electric field scattered by the $(m, n)^{\text{th}}$ element is given as \cite{29}
\begin{equation}
E_{mn}(\theta_\text{r},\phi_\text{r})
= E_0 \, \Gamma_{mn} \, f(\theta_\text{r},\phi_\text{r}) \,
e^{- \jmath k_0 (\hat{\mathbf{u}}_{\mathrm{inc}} \cdot \mathbf{p}_{mn})} \,
e^{\jmath k_0 (\hat{\mathbf{u}}_{\mathrm{r}} \cdot \mathbf{p}_{mn}) } \, ,
\label{eq:single_element_field}
\end{equation}
where the first exponential represents the phase of the incident wave at the element's location, and the second exponential accounts for the radiation toward the observation point, where $
\hat{\mathbf{u}}_{\mathrm{r}}
= \hat{\mathbf{x}} \sin\theta_{\text{r}} \cos \phi_{\text{r}}
+ \hat{\mathbf{y}} \sin\theta_{\text{r}} \sin \phi_{\text{r}}
+ \hat{\mathbf{z}} \cos \theta_{\text{r}} \, , $ Thus, the total reflected field considering all $M \times N$ RIS elements is obtained as
\begin{equation}
E(\theta_\text{r},\phi_\text{r})
= E_0 \sum_{m=1}^{M} \sum_{n=1}^{N}
\Gamma_{mn} \, f(\theta_\text{r},\phi_\text{r}) \,
e^{\jmath k_0 \left( \mathbf{p}_{mn} \cdot
\left(\hat{\mathbf{u}}_{\mathrm{r}} - \hat{\mathbf{u}}_{\mathrm{inc}} \right) \right)} .
\label{eq:ris_far_field_general}
\end{equation}

Furthermore, the directional element factor is approximated as
$ f(\theta_\text{r},\phi_\text{r}) \approx \cos^{q}(\theta_\text{r}) \, , $
where $q$ is an empirically chosen exponent controlling angular directivity. It is to be noted from (\ref{eq:ris_far_field_general}) that the beam steering is achieved by appropriately assigning the element reflection phases $\psi_{mn}$ defined in (\ref{refl}) across the RIS array.

\vspace{-0.05cm}
\section{Experimental Setup and Measurement Procedure}
\label{ssm}
\begin{figure}[!t]
	\centering   \includegraphics[width=1\columnwidth,height=8cm]{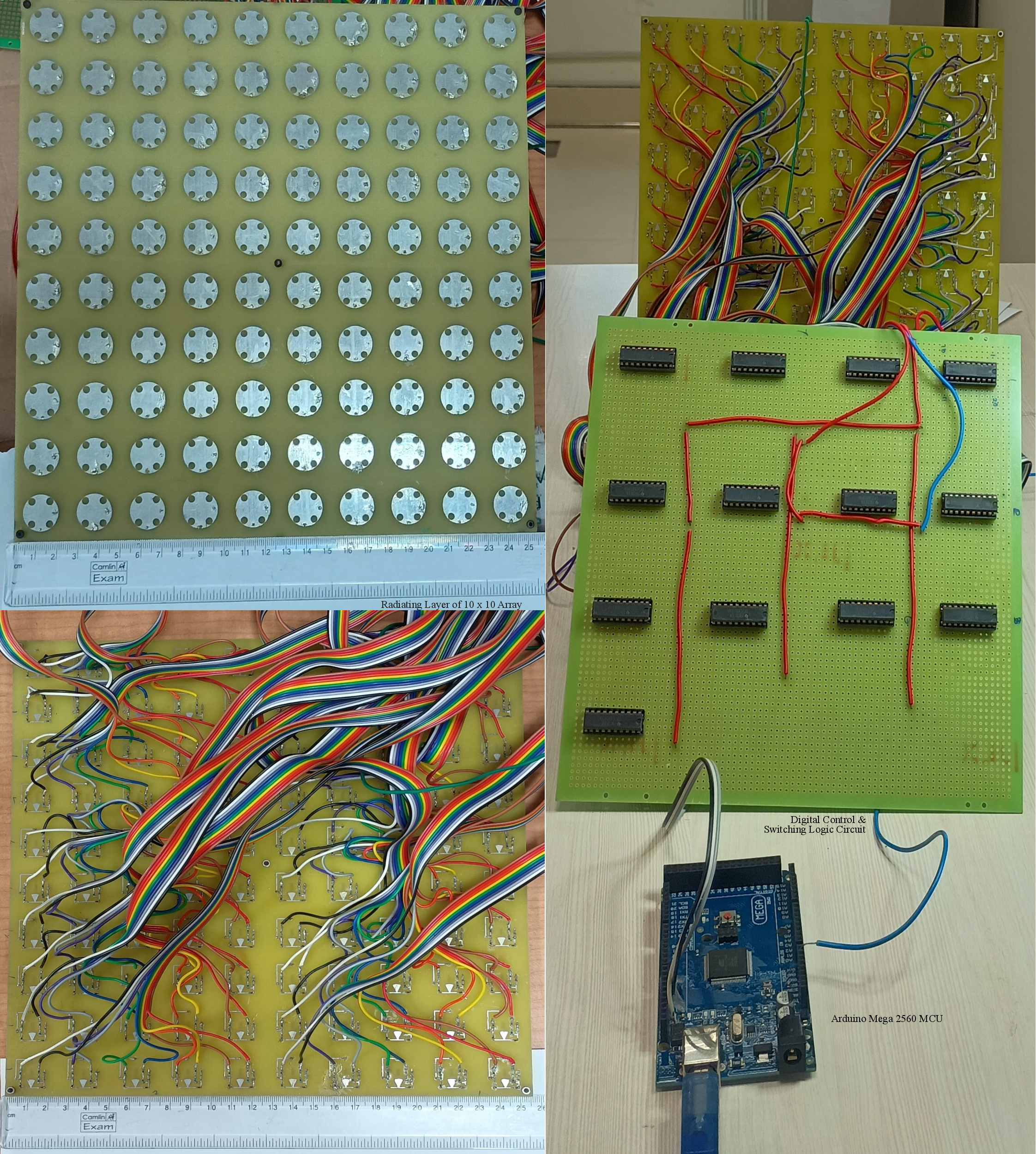}
	\caption{Photographs showing the RIS radiation layer array (top left), biasing layer (bottom left), and MCU-shift register-based control circuit driving the RIS (right).}
	\label{fig:pr2}
\end{figure}

Using the unit cell design, an RIS with 100 elements arranged in a $10\times10$ array is constructed. The array size is selected as a trade-off between beamforming gain, aperture size, and hardware complexity. Increasing the number of elements improves the array gain and angular resolution, but also increases the biasing and control complexity, while a smaller array limits the effective beam steering. Therefore, a $10\times10$ configuration is chosen. Figure~\ref{fig:pr2} shows the fabricated RIS hardware setup, including the radiation layer array and biasing network. Passive components and PIN diodes are integrated within each unit cell. The connections following Fig.~\ref{fig:rset} control architecture is based on a cascade of shift registers, which enables efficient handling of a large number of control lines while minimizing hardware complexity. An Arduino microcontroller unit having R3 ATmega2560 MCU \cite{mcu} sequentially programs the cascaded shift registers (TPIC6B595N, Texas Instruments) \cite{shiftregister} to individually configure the state of each unit cell across the array. A regulated DC power supply, controlled through the MCU-cum-shift register circuit, is used to bias the (Skyworks SMP1345-079LF) \cite{pindiode} PIN diodes in the RIS unit cells. The overall control scheme is shown in Fig.~\ref{fig:rset}. Measurements were performed by varying the Rx antenna in the azimuth plane from $-90^{\circ}$ to $+90^{\circ}$ with an interval of $5^{\circ}$ steps, recording the transmission coefficient magnitude ($S_{21}$) in dB for each steering angle. Furthermore, to achieve higher aperture efficiency, the focal-to-aperture ratio $(F/D)$ of the antenna is optimized, where $F$ is the distance between the feed antenna and the RIS, and $D$ denotes the length of the RIS aperture. In this work, $F/D=1.1$ is selected.
The $10\times10$ RIS array has an overall size of $255\,\text{mm} \times 255\,\text{mm}$, which corresponds to $2.9\lambda_g \times 2.9\lambda_g$ at $3.5\,\text{GHz}$ and operates at the center of the n78 band. 

\begin{figure}[!t]
	\centering
   \includegraphics[
    width=1\linewidth,
    height=6.5cm,
    keepaspectratio,
     trim=0mm 00mm 0mm 0mm,
    clip]{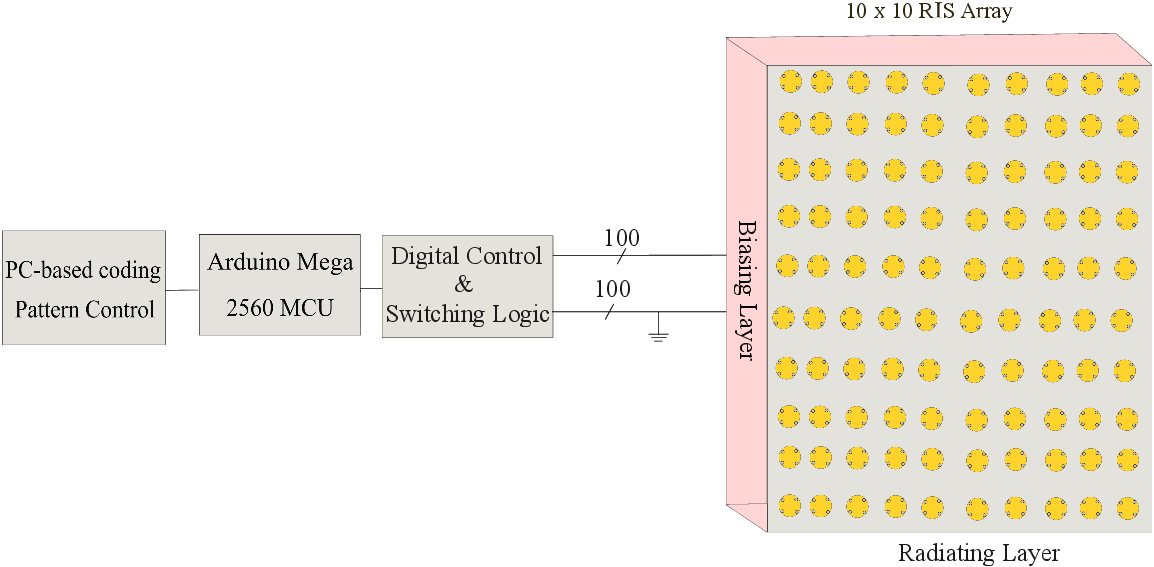} 
    \caption{Equivalent block diagram of the RIS setup in Fig. \ref{fig:pr2}.}
	\label{fig:rset}
\end{figure}

\begin{figure}[!t]
	\centering
	\includegraphics[width=1\columnwidth]{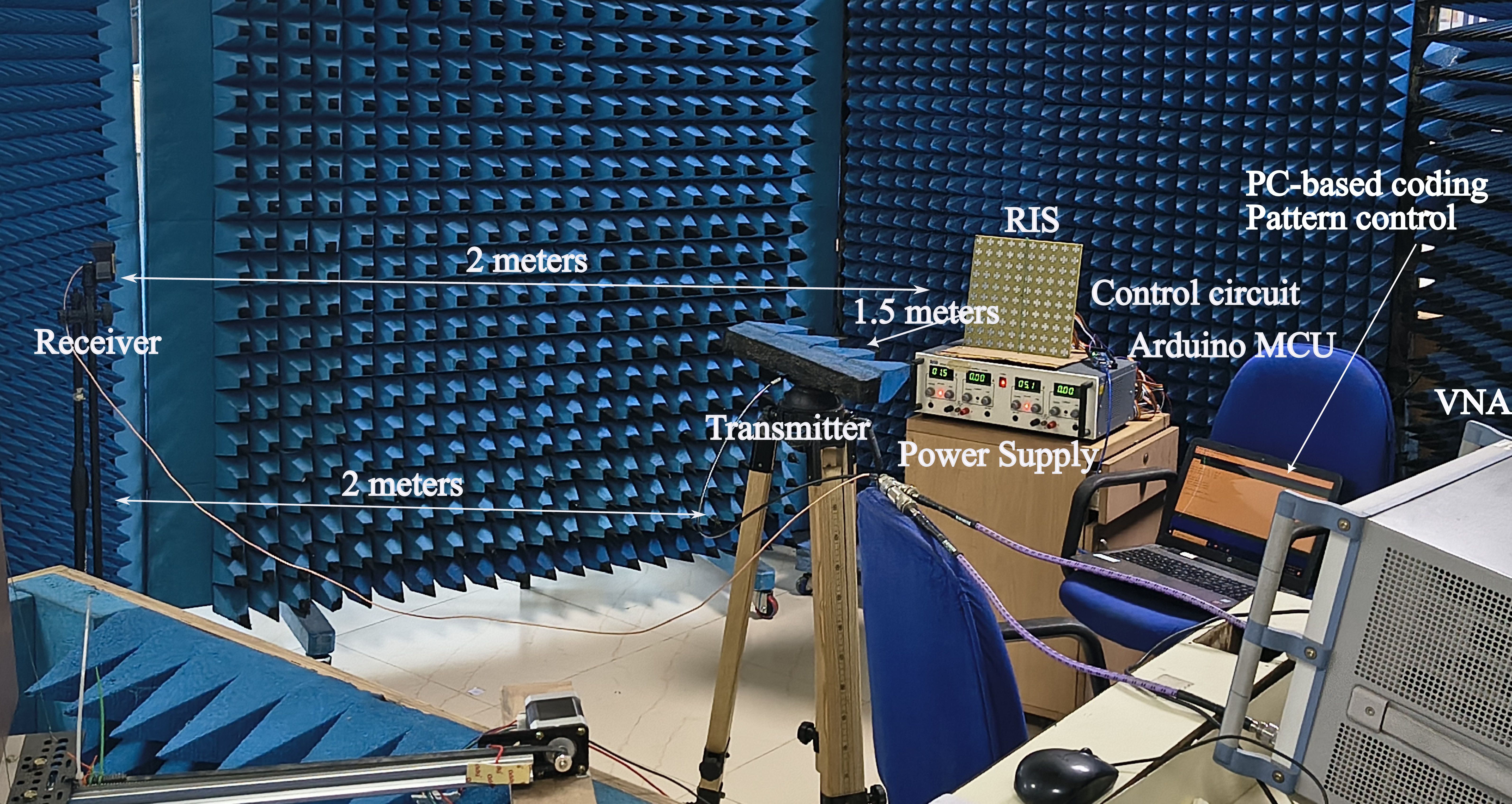}
	\caption{Complete measurement setup in anechoic chamber showing the Tx/Rx antennas, the proposed RIS with the PC-interfaced control circuit, with VNA and DC power supply.}
	\label{fig:pr1}
\end{figure}
The transmitting (Tx) and receiving (Rx) antennas are rectangular microstrip patch antennas (MPAs), with the RIS placed between them to assist signal transmission. The antennas are oriented to eliminate any direct line-of-sight (LOS) or side-lobe coupling, ensuring that the received signal originates only from RIS reflection. The complete measurement setup inside the anechoic chamber is shown in Fig.~\ref{fig:pr1}. During measurements, the Tx antenna is placed 1.5\,m from the RIS, while the Rx antenna is positioned 2\,m away along the reflected path, satisfying the far-field condition at the operating frequency. A vector network analyzer (VNA) is used to generate the transmit signal and to measure the received signal at the Rx antenna. The VNA output power is maintained at a constant level of 10\,dBm throughout the measurement process to ensure that the weakest received signal is sufficiently above the noise floor.

\vspace{-0.05cm}

\section{Experimental Verification and Results}
\label{svm}
The simulated response of the RIS is obtained from full-wave electromagnetic simulations and operates over 3.38–3.67\,GHz, with a center frequency of 3.5\,GHz. The experimental measurement is performed using a bistatic measurement setup, where the transmission coefficient magnitude $S_{21}$ is measured between the Tx and Rx MPAs in the presence of the RIS. Both antennas operate between 3-4\,GHz, with a center frequency of 3.5\,GHz and provide a gain of about 6.28dBi.  
To compare, we initially used a normal copper reflector plate and obtained measured $S_{21}$ of -54.2\,dB. Thereafter, replacing it with the proposed RIS showed an improved $S_{21}$ of -45.2\,dB. The incident wave from the transmitting antenna was kept at normal incidence, $(\theta_\text{inc}, \phi_\text{inc}) = (0^\circ, 0^\circ)$, while the receiving antenna was rotated horizontally to capture the reflected beam. The simulated and measured reflection directions are summarized in Table \ref{s00}.
\begin{table}[t]
\centering
\caption{Simulated and measured beam steering of the proposed RIS for MPA illumination at $(\theta_\text{inc}, \phi_\text{inc}) = (0^\circ, 0^\circ)$.}
\label{s00}
\setlength{\tabcolsep}{12pt}
\renewcommand{\arraystretch}{1.3}

\begin{tabular}{cccc}
\toprule \toprule
Simulated & Simulated & Measured & Measured \\
$(\theta_r,\phi_r)$ & Gain (dBi) & $(\theta_r,\phi_r)$ & $S_{21}$ (dB) \\
\midrule
$(0^\circ,0^\circ)$ & 10.2 & $(0^\circ,4^\circ)$ & -45.2 \\
$(0^\circ,-17^\circ)$ & 9.9 & $(0^\circ,-10^\circ)$ & -49.0 \\
$(0^\circ,22^\circ)$ & 9.6 & $(0^\circ,26^\circ)$ & -55.3 \\
$(0^\circ,28^\circ)$ & 9.5 & $(0^\circ,32^\circ)$ & -55.5 \\
$(0^\circ,30^\circ)$ & 9.48 & $(0^\circ,31^\circ)$ & -58.3 \\
$(0^\circ,-30^\circ)$ & 9.41 & $(0^\circ,-27^\circ)$ & -58.5 \\
\bottomrule
\bottomrule
\end{tabular}
\end{table}

\begin{figure}[t]
\centering

\subfloat[]{
\includegraphics[width=0.46\columnwidth]{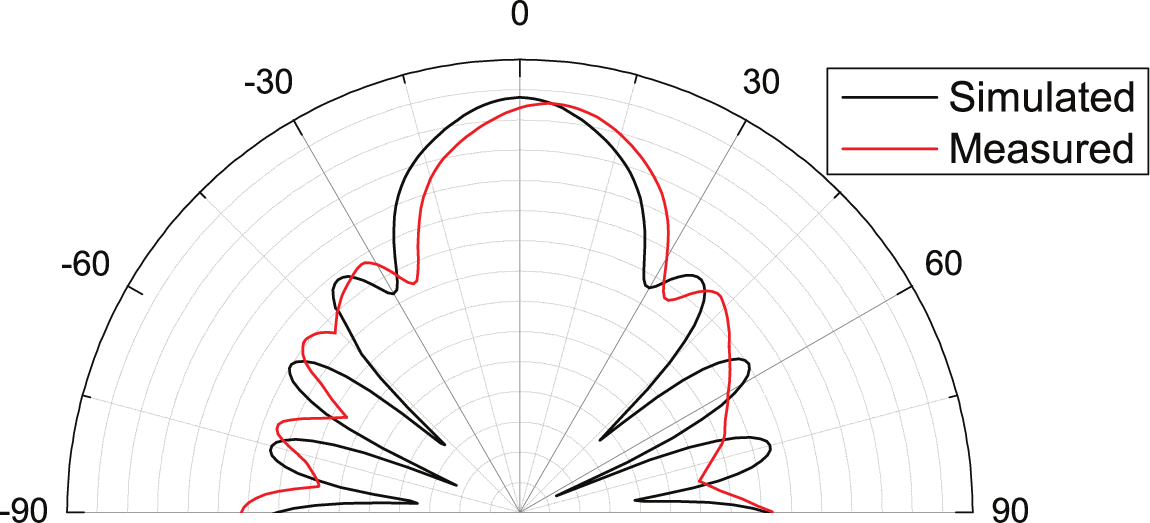}
}\hfill
\subfloat[]{
\includegraphics[width=0.46\columnwidth]{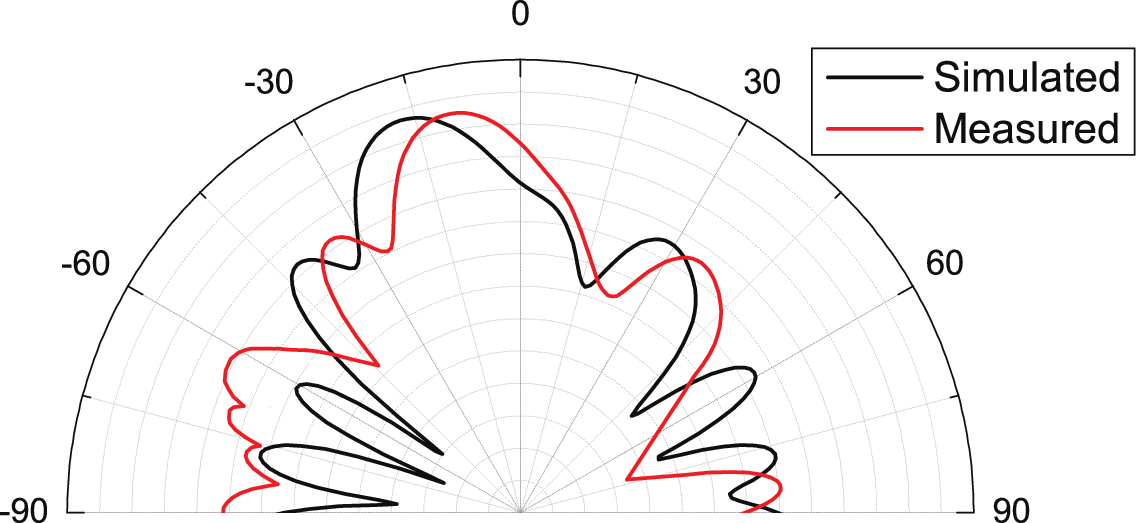}
}

\vspace{4pt}

\subfloat[]{
\includegraphics[width=0.46\columnwidth]{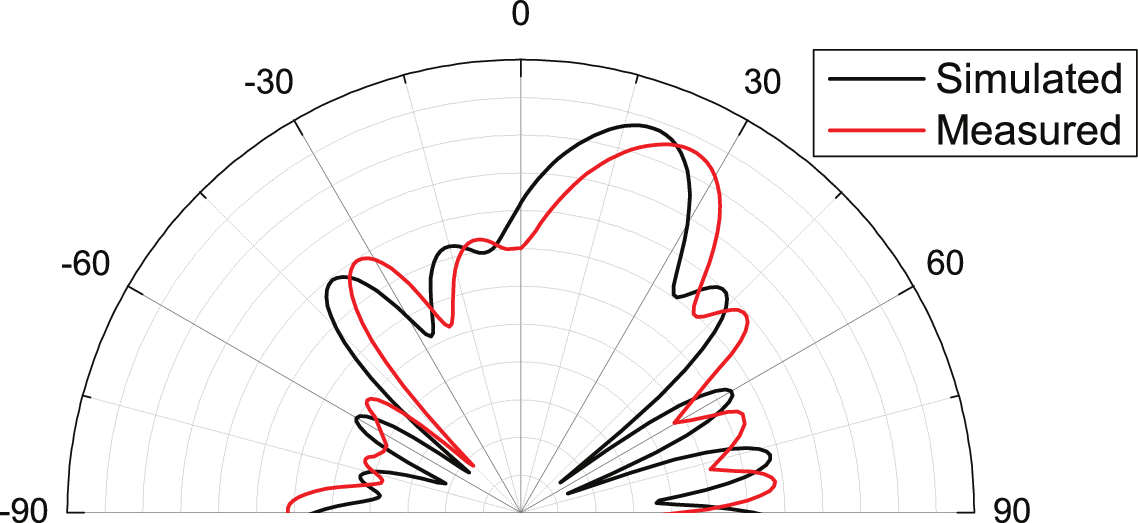}
}\hfill
\subfloat[]{
\includegraphics[width=0.46\columnwidth]{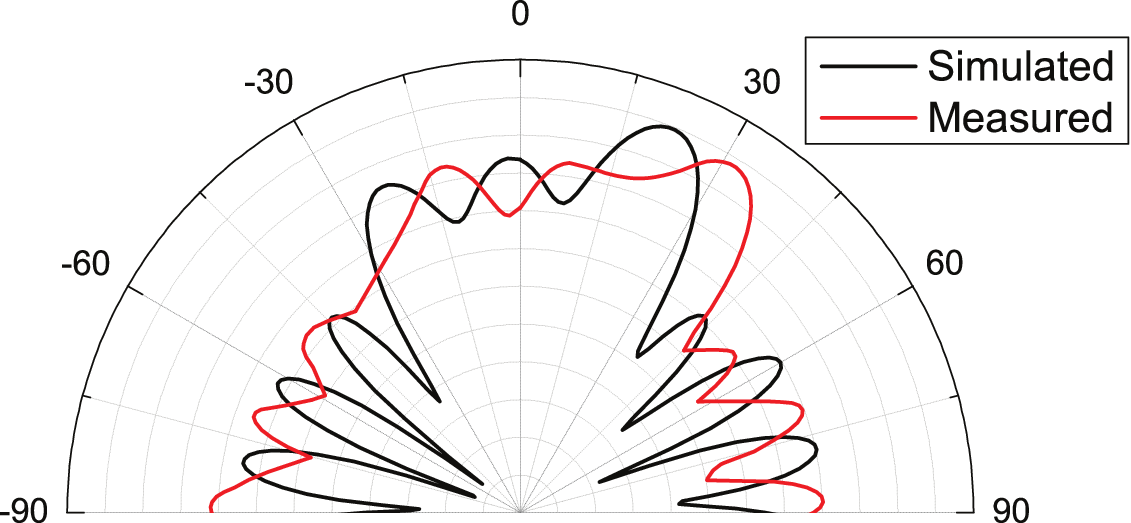}
}\vspace{4pt}

\subfloat[]{
\includegraphics[width=0.46\columnwidth]{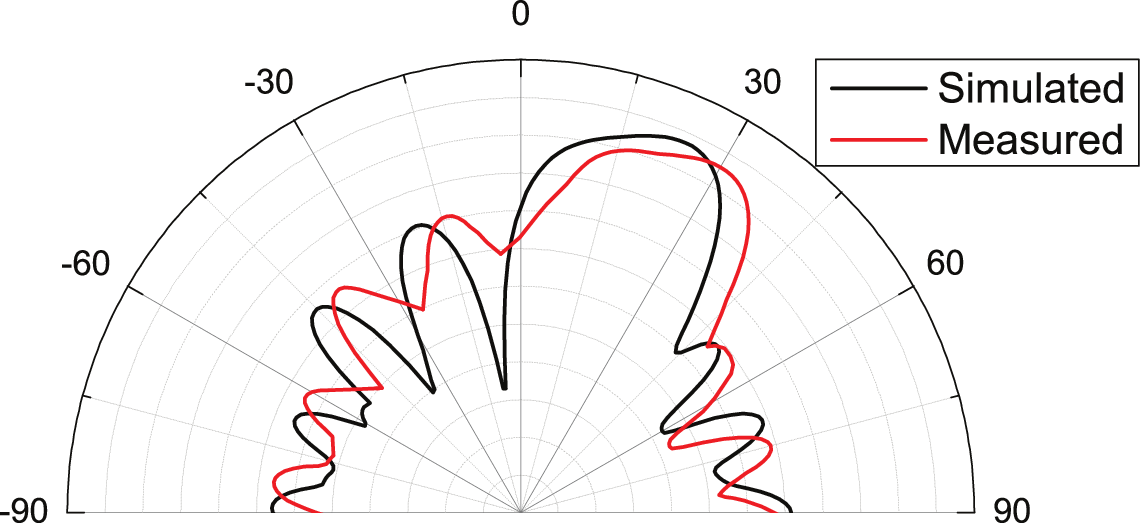}
}\hfill
\subfloat[]{
\includegraphics[width=0.46\columnwidth]{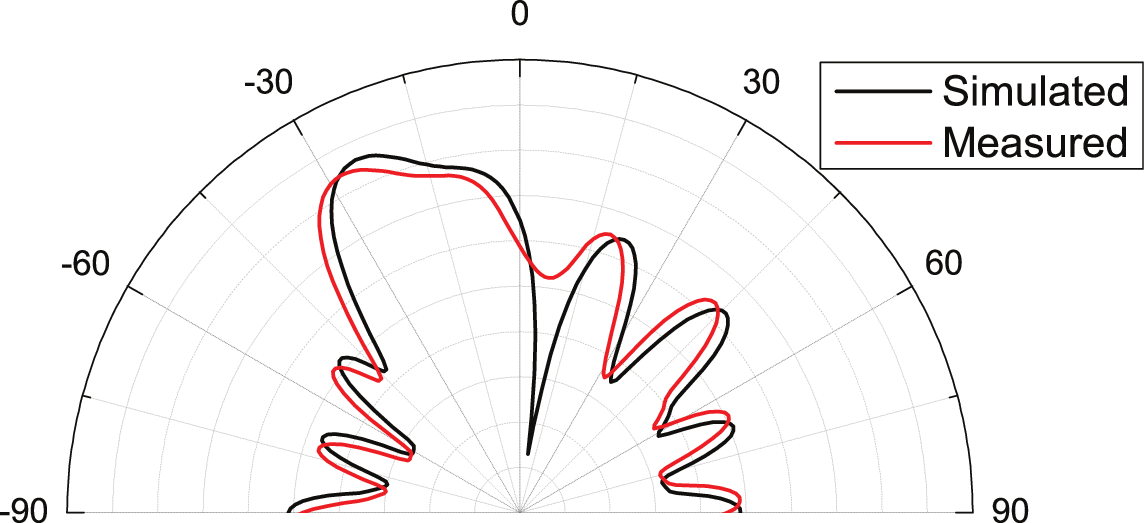}
}

\caption{Comparison of simulated and measured normalized radiation patterns for steering angles of (a) $0^\circ$, (b) $-17^\circ$, (c) $22^\circ$, (d) $28^\circ$, (e) $30^\circ$, and (f) $-30^\circ$.}

\label{rp}
\end{figure}

\begin{figure}[ht]
\centering

\subfloat[]{
\includegraphics[width=0.3\columnwidth]{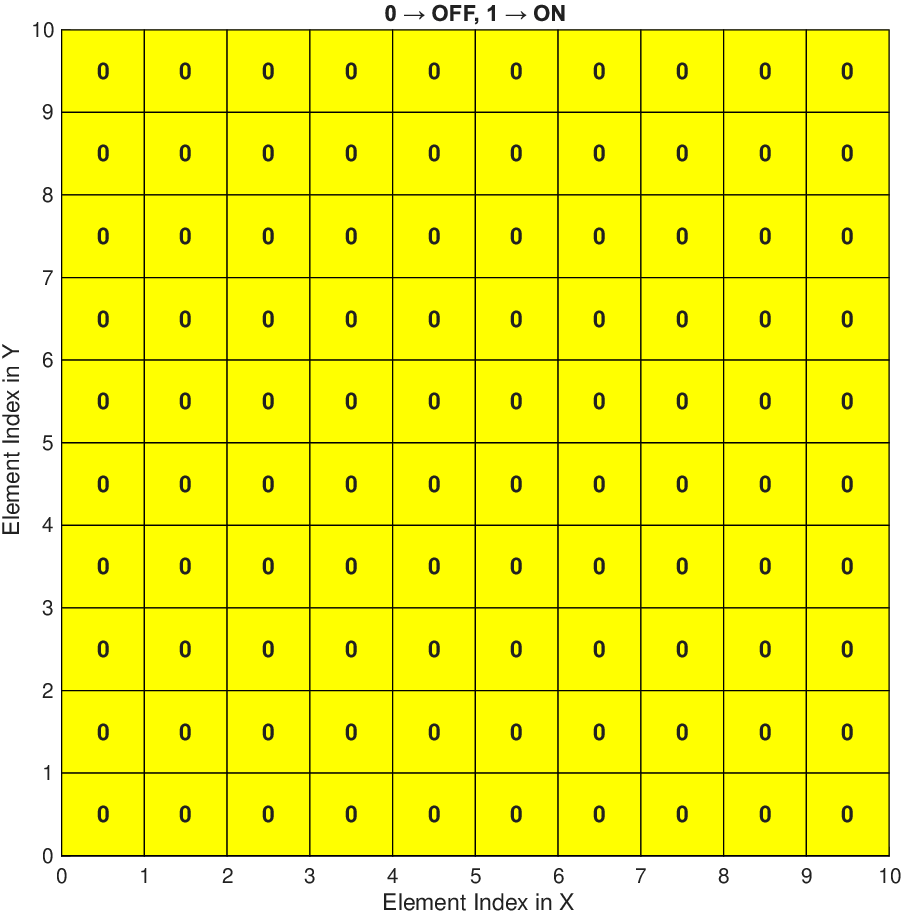}
}\hfill
\subfloat[]{
\includegraphics[width=0.3\columnwidth]{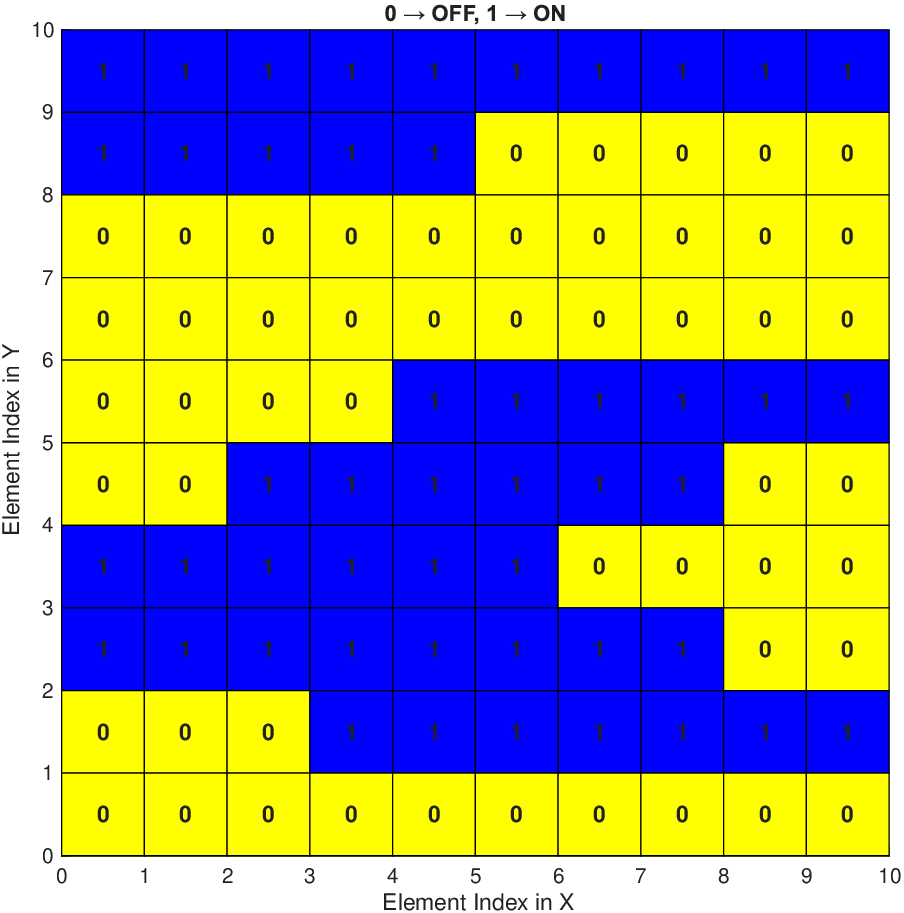}
}\hfill
\subfloat[]{
\includegraphics[width=0.3\columnwidth]{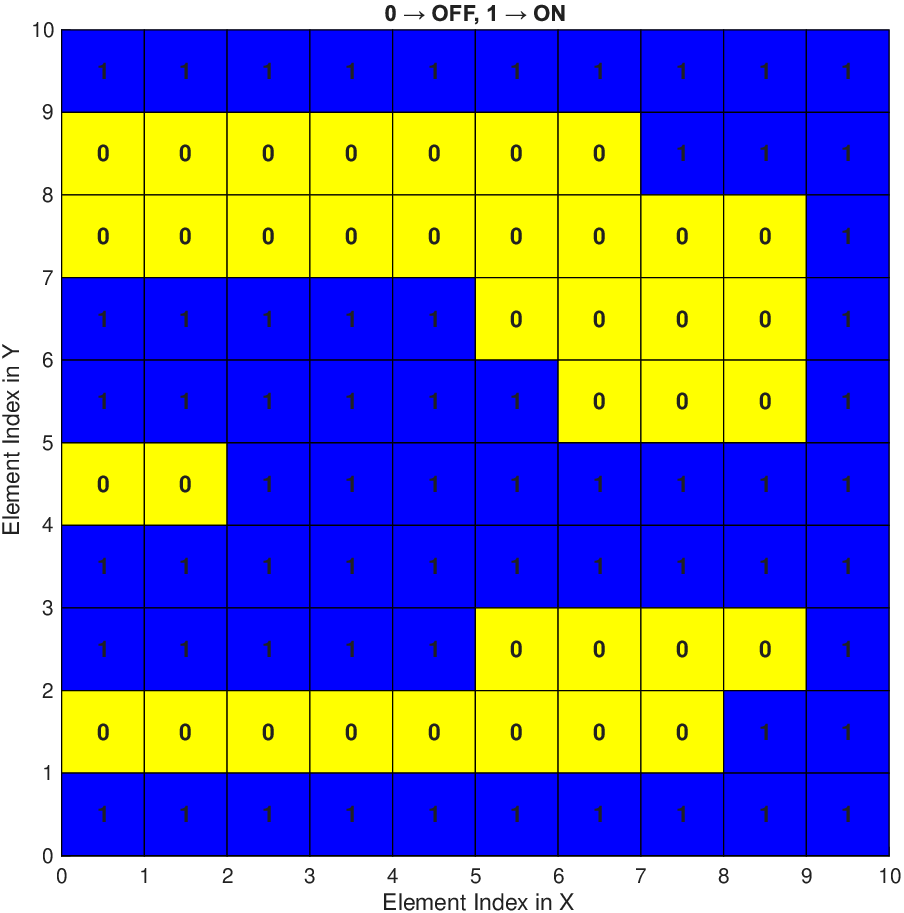}
}

\vspace{4pt}

\subfloat[]{
\includegraphics[width=0.3\columnwidth]{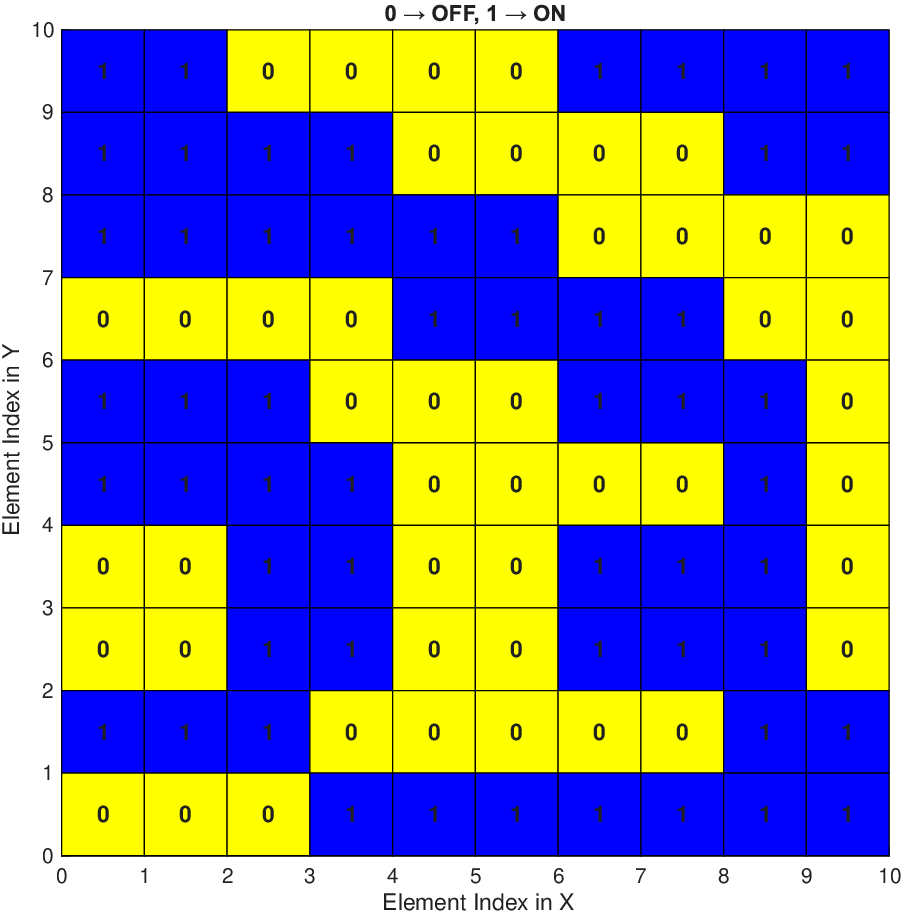}
}\hfill
\subfloat[]{
\includegraphics[width=0.3\columnwidth]{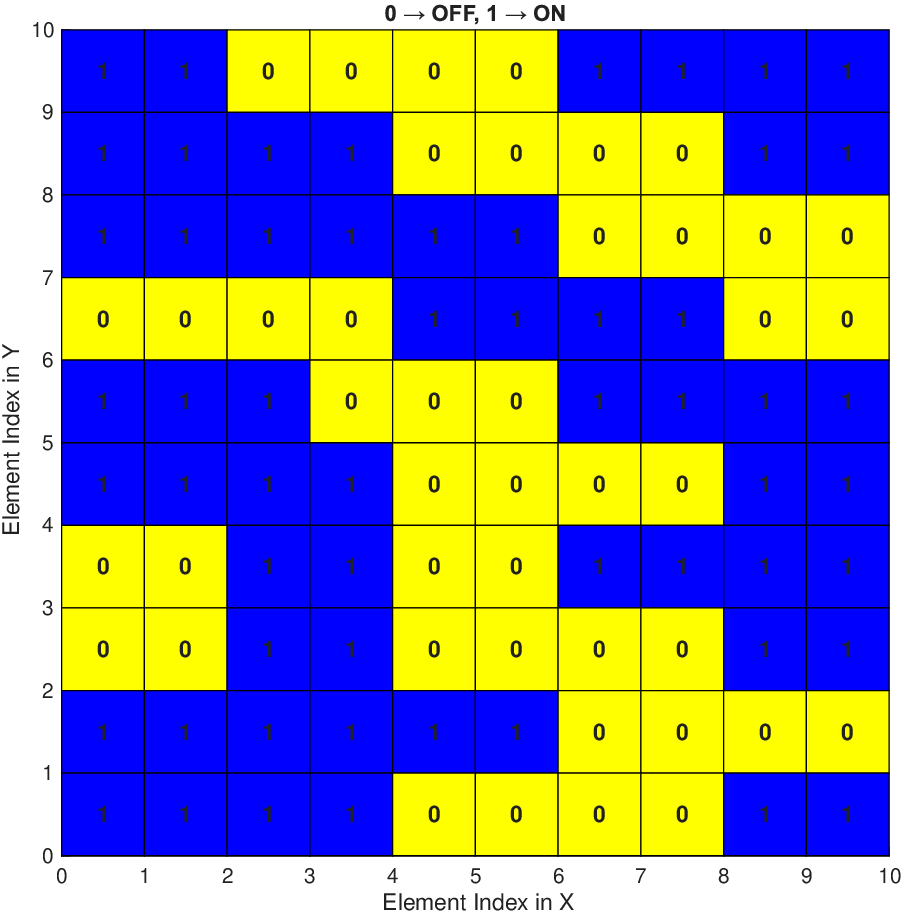}
}\hfill
\subfloat[]{
\includegraphics[width=0.3\columnwidth]{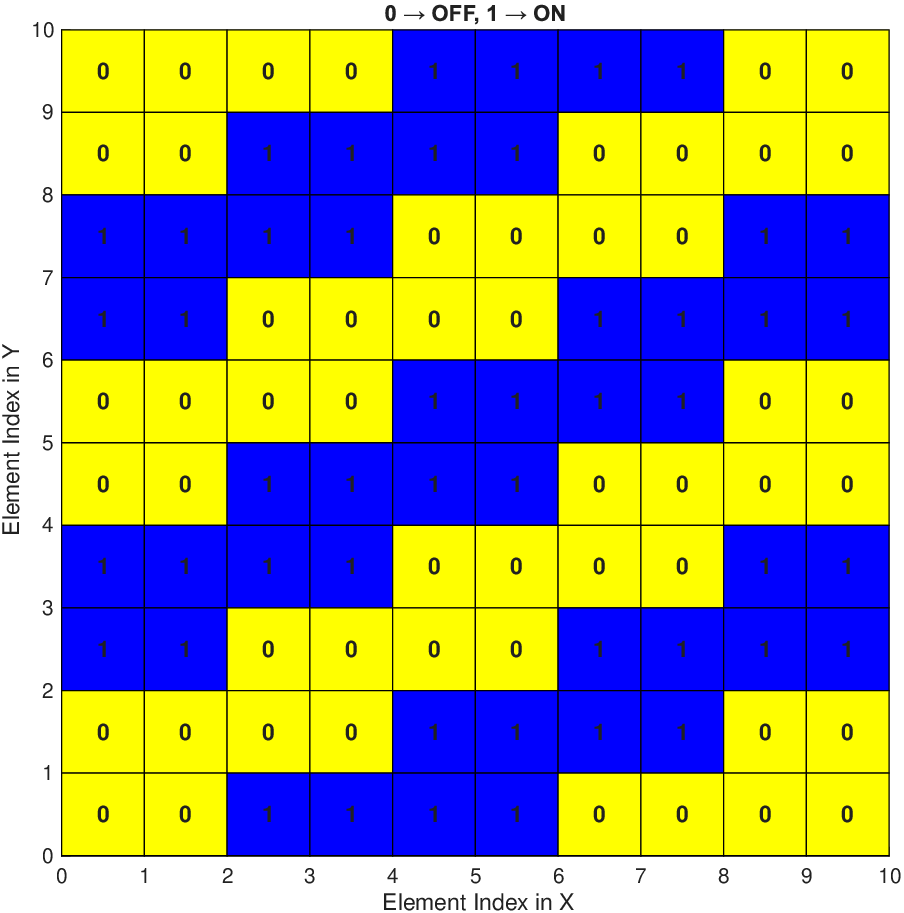}
}

\caption{Coding patterns applied to the RIS for steering angles of (a) $0^\circ$, (b) $-17^\circ$, (c) $22^\circ$, (d) $28^\circ$, (e) $30^\circ$, and (f) $-30^\circ$.}
\label{fcp}
\end{figure}

The close agreement between simulated and measured directions demonstrates that the RIS successfully steers the reflected beam close to the intended directions.
Since the measured $S_{21}$ using the VNA includes free-space path loss, cable losses, antenna mismatch, and hardware attenuation, a direct comparison with absolute simulated gain is not appropriate. Therefore, both simulated and measured responses are normalized to their respective maximum values to enable a comparison of angular pattern shapes. The normalized simulated angular response is obtained from the far-field data and is expressed as
\begin{equation}
G_{\mathrm{norm}}(\theta) = G(\theta) - \max_{\theta}\{G(\theta)\},
\end{equation}
where $G(\theta)$ represents the simulated gain in dB as a function of observation angle $\theta$. Similarly, the measured response is normalized using
\begin{equation}
S_{21,\mathrm{norm}}(\theta) = S_{21}(\theta) - \max_{\theta}\{S_{21}(\theta)\},
\end{equation}
where $S_{21}(\theta)$ denotes the measured transmission coefficient in dB. After normalization, both simulated and measured patterns become relative and dimensionless, allowing direct comparison of beam direction and shape. The radiation patterns corresponding to different steering angles are shown in Fig.~\ref{rp}, with their respective coding configurations illustrated in Fig.~\ref{fcp}. The coding patterns shown in Fig.~\ref{fcp} are generated using the RIS beam steering formulation summarized in Section \ref{stm}. 

\begin{figure}

	    \centering
	    \includegraphics[width=\linewidth,trim=10 20 140 50,clip]{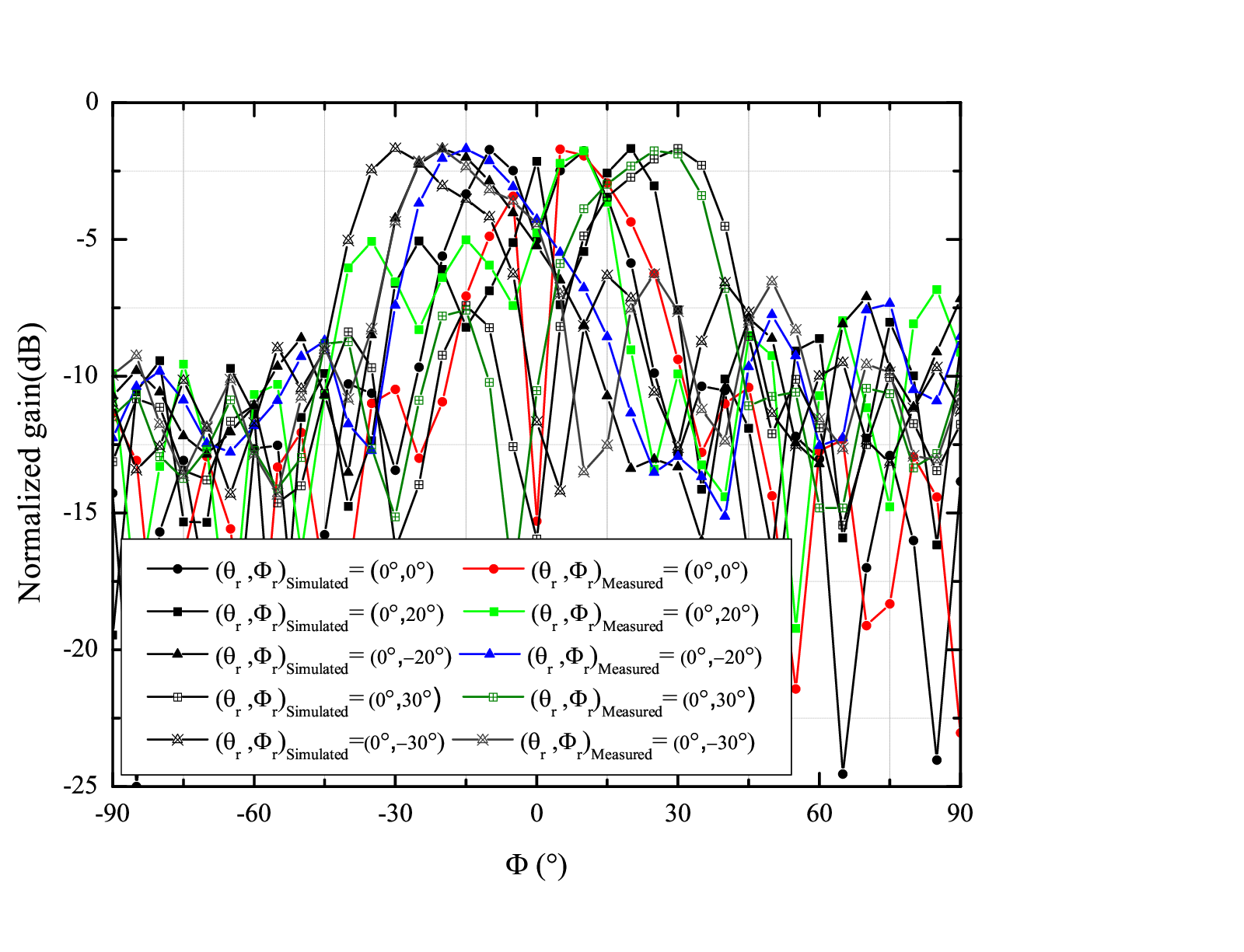}

	 \caption{Normalized radiation patterns of the proposed RIS for an incident angle of $(\theta_{inc},\phi_{inc})=(0^\circ,30^\circ)$. The reflected beam direction is observed for different  azimuth angles.}
	    \label{fig:nc1}
\end{figure}
\begin{figure}

	    \centering
	    \includegraphics[width=\linewidth,trim=60 20 90 60,clip]{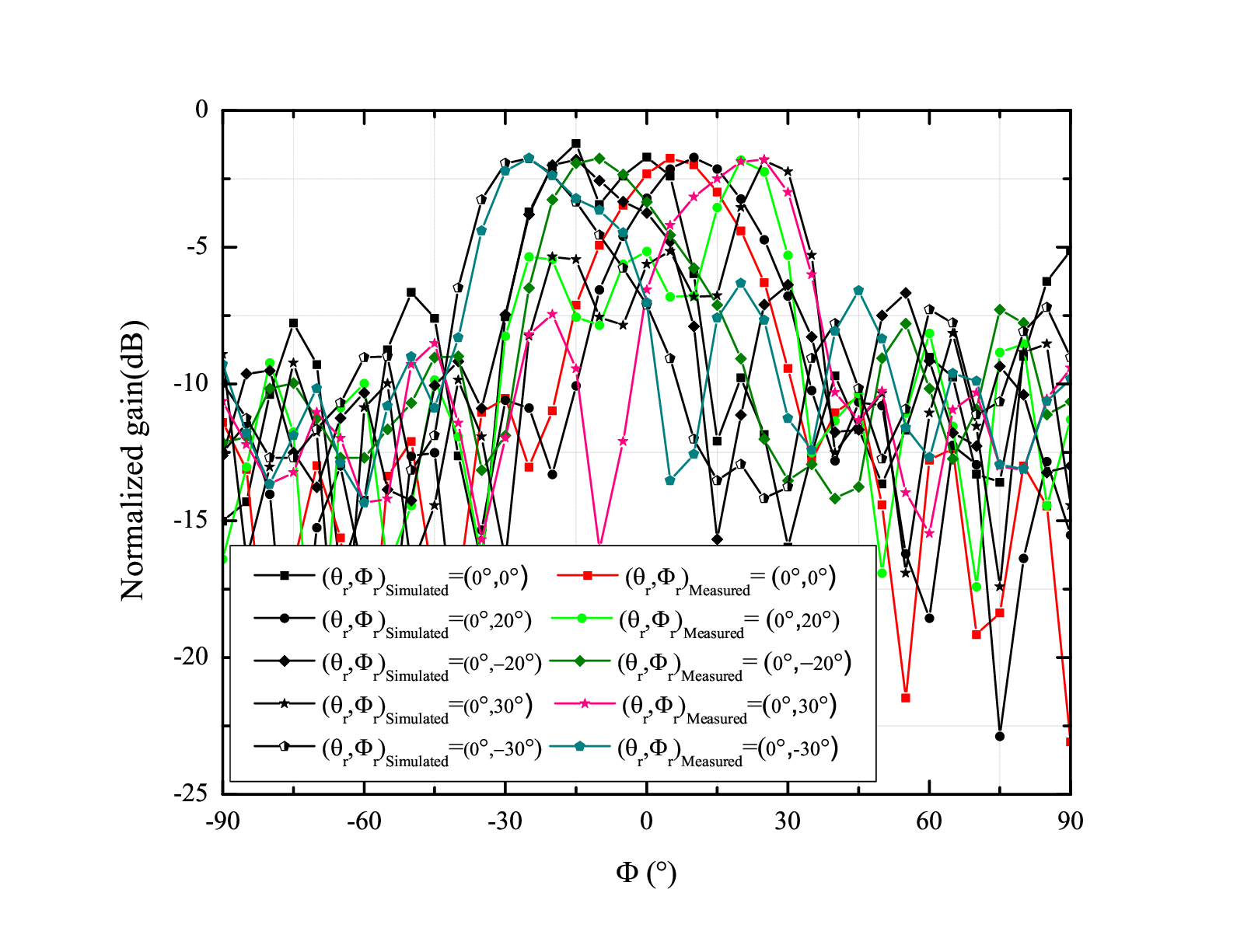}

	   \caption{Normalized radiation patterns of the proposed RIS for an incident angle of $(\theta_{inc},\phi_{inc})=(0^\circ,-30^\circ)$. The reflected beam direction is observed for different azimuth angles.}
	    \label{fig:nc2}
\end{figure}

Subsequently in the experimental setup, the Tx antenna incident angle was fixed at an elevation angle of $(\theta_{\text{inc}}, \phi_{\text{inc}}) = (0^\circ,\,30^\circ)$,  and later at $(\theta_{\text{inc}}, \phi_{\text{inc}}) = (0^\circ,\,\text{-}30^\circ)$, while the Rx antenna was swept horizontally to capture the reflected beam. The received signal magnitude ($S_{21}$) was recorded at 5° intervals to capture the full angular response, as shown in Fig.~\ref{fig:nc1}, when the incident angle is set to $(\theta_{inc},\phi_{inc})=(0^\circ,30^\circ)$. The simulated and measured reflection directions are summarized in Table \ref{s01}.
\begin{table}[!t]
\centering
\caption{Simulated and measured beam steering by the proposed RIS for MPA illumination at $(\theta_\text{inc}, \phi_\text{inc}) = (0^\circ, 30^\circ)$.}
\label{s01}
\setlength{\tabcolsep}{12pt}
\renewcommand{\arraystretch}{1.3}
\begin{tabular}{cccc}
\toprule \toprule
Simulated & Simulated & Measured & Measured \\
$(\theta_r,\phi_r)$ & Gain (dBi) & $(\theta_r,\phi_r)$ & $S_{21}$ (dB) \\
\midrule
$(0^\circ,0^\circ)$ & 8.4 & $(0^\circ,-4^\circ)$ & -54.8 \\
$(0^\circ,20^\circ)$ & 7.9 & $(0^\circ,26^\circ)$ & -56.5 \\
$(0^\circ,30^\circ)$ & 7.6 & $(0^\circ,27^\circ)$ & -58.2 \\
$(0^\circ,-20^\circ)$ & 9.9 & $(0^\circ,-18^\circ)$ & -46.0 \\
$(0^\circ,-30^\circ)$ & 9.5 & $(0^\circ,-33^\circ)$ & -46.3 \\
\bottomrule
\bottomrule
\end{tabular}
\end{table}
From Fig.~\ref{fig:nc2}, when the incident angle is set to $(\theta_{inc},\phi_{inc})=(0^\circ,\text{-}30^\circ)$ and the Rx antenna is set with a different Rx angle, the simulated and measured reflection directions are summarized in Table \ref{s02}. Overall, there is a good agreement between the simulated and measured results.

\begin{table}[t]
\centering
\caption{Simulated and Measured Beam Steering by the proposed RIS for MPA illumination at $(\theta_{inc},\phi_{inc})=(0^\circ,-30^\circ)$}
\label{s02}
\setlength{\tabcolsep}{12pt}
\renewcommand{\arraystretch}{1.3}

\begin{tabular}{cccc}
\toprule \toprule
Simulated & Simulated & Measured & Measured \\
$(\theta_r,\phi_r)$ & Gain (dBi) & $(\theta_r,\phi_r)$ & $S_{21}$ (dB) \\
\midrule
$(0^\circ,5^\circ)$ & 7.8 & $(0^\circ,6^\circ)$ & -49.1 \\
$(0^\circ,22^\circ)$ & 9.1 & $(0^\circ,24^\circ)$ & -52.3 \\
$(0^\circ,26^\circ)$ & 9.4 & $(0^\circ,25^\circ)$ & -45.8 \\
$(0^\circ,-18^\circ)$ & 7.8 & $(0^\circ,-19^\circ)$ & -58.0 \\
$(0^\circ,-25^\circ)$ & 7.6 & $(0^\circ,-28^\circ)$ & -56.5 \\
\bottomrule
\bottomrule
\end{tabular}
\end{table}

\begin{table*}[!t]
\centering
\caption{Comparison of the proposed RIS with state of the art RIS hardware.}
\label{tab:comparison}
\renewcommand{\arraystretch}{2.5}

\resizebox{\textwidth}{!}{
\begin{tabular}{lccccccccccc}
\toprule\toprule
Ref. & $(\varepsilon_\text{r})$ &
Aperture Size &
Unit Cell Tuning &
Beam Steering &
Freq. (GHz) &
Control Method &
Biasing Network &
$F/D$ &
Control Lines  \\
\midrule

{\cite{23}} & 2.2  & $7.9\lambda_g \times 7.9\lambda_g$ & 1-bit (Varactor Diode) & $\pm40^\circ$ & 7.45 & FPGA-based & Inductor-based & 0.5 & 400 \\

{\cite{24}} & 3.0 & $8\lambda_g \times 8\lambda_g$ & 1-bit / quasi-2-bit(PIN Diode) & $\pm60^\circ$ & 5 & D flip-flop–based & Integrated with structure  & 1 &20 \\

{\cite{25}} & 3.0 & $8\lambda_g \times 8\lambda_g$ & 2-bit (PIN Diode)  & $\pm60^\circ$ & 5 & FPGA-based &  single-layer folded ground & 1  & 128  \\

{\cite{26}} & 4.4 & $10\lambda_g \times 10\lambda_g$ & 1-bit and 2-bit digital coding & $\pm30^\circ$ & 5.9 & Passive coded based  & Not required  & -- & Passive  \\

\textbf{This Work} & \textbf{4.4} & \textbf{$2.9\lambda_g \times 2.9\lambda_g$} & \textbf{1-bit(PIN Diode)}  & \textbf{$\pm30^\circ$} & \textbf{3.5} & \textbf{MCU-based} & \makecell{\textbf{Compact RF-DC }\\ \textbf{decoupling network}} &   \textbf{1.1}  & \textbf{100}  \\

\bottomrule\bottomrule
\end{tabular}}
\end{table*}

Table \ref{tab:comparison} compares the proposed RIS prototype with previously reported RIS hardware implementations in terms of substrate dielectric constant, aperture size, tuning mechanism, beam steering capability, operating frequency, control method, biasing network, the focal-to-aperture ratio $(F/D)$, and number of control lines. Most existing designs rely on FPGA-based control and more complex biasing networks, which increase hardware complexity. In contrast, the proposed design in this paper uses a simpler MCU-based control with a more simplified bias network. While keeping the aperture size compact, the proposed RIS still enables practical beam steering using a 1-bit PIN-diode configuration.

\vspace{-0.05cm}
\section{Effect of the RIS on QPSK Constellation}
\label{gnuradio}
\begin{figure}
    \centering
    \includegraphics[height=4.5cm,width=1\linewidth]{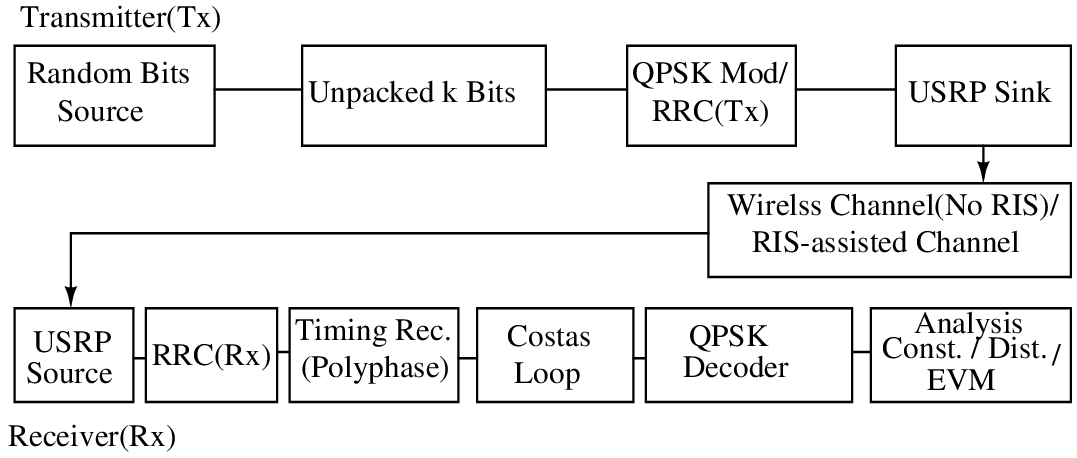}
    \caption{Block diagram of the USRP-based QPSK transmission
and reception chain with RIS integration
}
    \label{fig:flow}
\end{figure}

Based on the antenna gain and the Tx–RIS–Rx link distances, QPSK is chosen as a suitable modulation scheme for reliable communication. RIS-assisted QPSK transmission is also tested experimentally in a noisy (non-anechoic) environment. To this end, we transmitted symbols using a NI USRP-2901\cite{usrp} platform controlled through GNU Radio with the center frequency of 3.5\,GHz. Each transmitted symbol is represented as
\begin{equation}
s_k = \sqrt{E_s}\,(a_k + \jmath b_k), \quad a_k, b_k \in \{\pm 1\},
\end{equation}
where $E_s$ denotes the symbol energy. The baseband processing chain implemented in GNU Radio, including symbol mapping, pulse shaping, and up-conversion, is illustrated in the block diagram in Fig.~\ref{fig:flow}. Without the RIS, the received signal at the antenna can be modeled as
\begin{equation}
r_k = h_{\text{ch}} s_k + n_k,
\end{equation}
where $h_{\text{ch}} = |h| e^{j \theta_h}$ represents the complex channel gain and $n_k$ denotes additive noise. The observed constellation at the receiver reflects the combined effects of amplitude scaling and phase rotation caused by $h_{\text{ch}}$. Depending on the magnitude $|h|$ and phase $\theta_h$, the constellation points may shrink or rotate, degrading symbol detection accuracy.

When the RIS is introduced with $M \times N$ programmable elements, each applying a phase shift $\psi_{mn}$, the effective channel is expressed as \cite{b22}
\begin{equation}
h_{\text{eff}} = h_\text{d} + \sum_{m=1}^{M} \sum_{n=1}^N h_{t,(m,n)} \, e^{\jmath \psi_{mn}} \, h_{r,(m,n)} \, ,
\end{equation}
where $h_\text{d}$ is the direct Tx--Rx channel, and $h_{t,(m,n)}$ and $h_{r,(m,n)}$ denote the Tx--RIS and RIS--Rx channel components for the $(m,n)^{\text{th}}$ element. By carefully tuning the phases $\psi_{mn}$, the magnitude of the effective channel $|h_{\text{eff}}|$ increases, enhancing the Euclidean separation between constellation points, while phase alignment minimizes rotation errors. After equalization, the received symbol estimate can be written as
\begin{equation}
\hat{s}_k = \frac{r_k}{h_{\text{eff}}} = s_k + \frac{n_k}{h_{\text{eff}}},
\label{eq:equalized_symbol}
\end{equation}
indicating that the variance of the post-equalization noise is inversely proportional to $|h_{\text{eff}}|^2$. Consequently, a larger $|h_{\text{eff}}|$, achieved through RIS phase optimization, reduces the variance of the post-equalization noise and tightens the symbol clusters in the $I–Q$ plane. The minimum distance between constellation points is defined as  \cite{28}
\begin{equation}
d_{\min} = \min_{\substack{i,j \\ i \neq j}} \left\| \hat{s}_i - \hat{s}_j \right\|_2,
\label{eq:dmin}
\end{equation}
where $\hat{s}_i$ and $\hat{s}_j$ are equalized received symbols. This metric directly depends on the effective channel: $
d_{\min}^{\text{const}} \propto |h_{\text{eff}}|.$

\begin{figure}[!t]
    \centering
    \includegraphics[width=\linewidth, trim=0 0 15mm 0, clip]{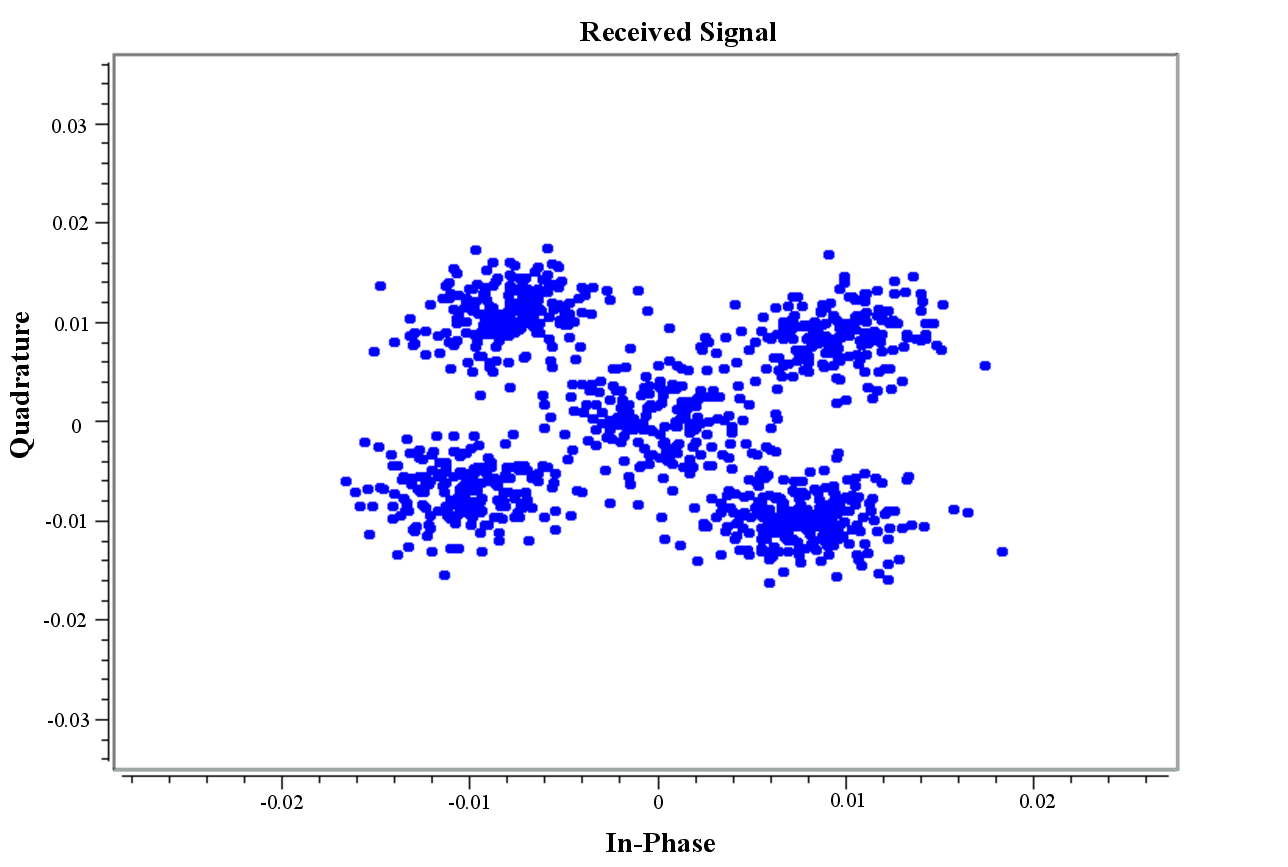 }  
    \caption{Constellation of received QPSK signal measured without RIS integration using a GNU Radio setup and USRP 2901.}
    \label{nris}
\end{figure}

Fig.~\ref{nris} shows the received signal without RIS assistance, including the QPSK constellation at the transmitter and the corresponding receiver signal. In this case, the channel loss and noise distort the received constellation, producing compressed symbol clusters. Consequently, mismatches appear between the transmitted (source) bits and the recovered (decoded) bits, indicating reduced link reliability.

\begin{figure}[!t]
    \centering
    \includegraphics[width=\linewidth, trim=0 0 15mm 0, clip]{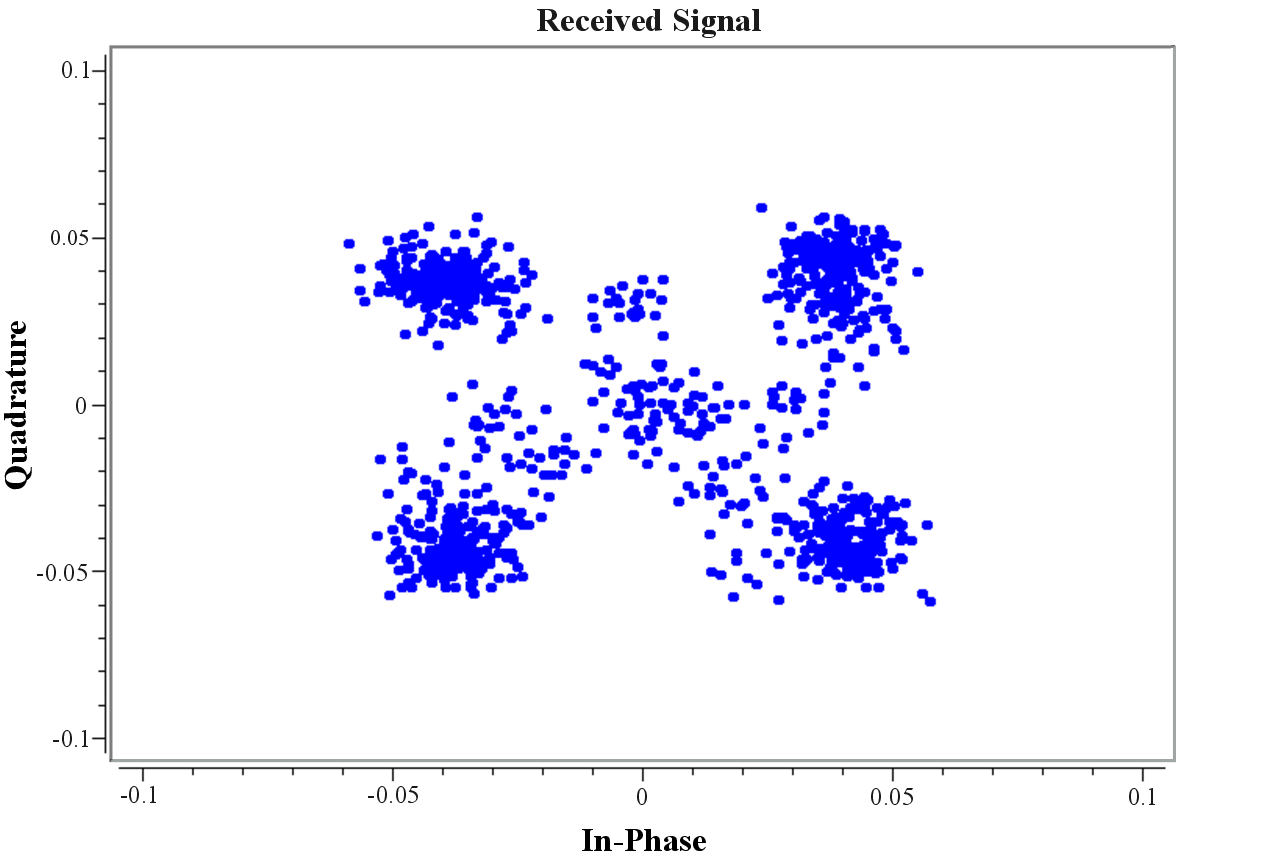}  
  \caption{Constellation of received QPSK signal measured with RIS integration using a GNU Radio setup and USRP 2901 in the same environment as Fig. \ref{nris}.}
    \label{wris}
\end{figure}

Fig.~\ref{wris} shows the measurements with the RIS enabled using a fixed phase configuration. Compared to Fig.~\ref{nris}, the received constellation becomes more compact and better separated, indicating improved channel conditions. Consequently, the decoded time-domain bits have fewer errors and better agreement with the source bits. These results confirm that RIS assistance improves the received signal quality without modifying the baseband processing.

\vspace{-0.05cm}

\section{Conclusion}
\label{sc}
In this paper, we demonstrate a 1-bit RIS that achieves beam steering up to $\pm30^\circ$ over the frequency range 3.38--3.67\,GHz, while utilizing a low-cost FR4 substrate. The inherent dielectric loss of the FR4 is alleviated by introducing an air gap between manually stacked up PCB layers, where the detailed procedure to calculate the ideal air gap using FOM and optimal via location(s) resulting in a new design for the unit cells, is also proposed. The biasing and control circuit for this RIS consists of PIN-diode-based circuits, controlled through an MCU and shift registers, thereby making this comparatively simpler as compared to other reported works. The proposed RIS achieves the $\pm30^\circ$ beam steering with an average phase deviation of $\pm3.75^\circ$ compared to the simulated results, showing good agreement. This is achieved by a new phase-gradient-based coding scheme, which is proposed and explained. For an incident wave angular range of $\pm30^\circ$, the RIS offers a main-lobe gain improvement of about 9\,dB, all while maintaining a compact aperture size of $2.9\lambda_g\times2.9\lambda_g$ $(255\times255\,\text{mm}^2)$. Furthermore, the RIS has also been tested in a noisy environment to transmit QPSK signals using an SDR, and a well-spaced symbol cluster in the received constellation is observed.

\FloatBarrier
\vspace{2mm}

 \bibliographystyle{IEEEtran}
 \bibliography{IEEEabrv,bibliography}

\end{document}